\documentclass[10pt, twoside]{article}

\usepackage{williams_notations}
\usepackage{williams_packages}
\usepackage{mathtools}
\usepackage{relsize}
\usepackage{algorithm}
\usepackage{algorithmic}
\usepackage{verbatim}
\usepackage{url}
\usepackage{hyperref}

\allowdisplaybreaks

\usepackage{lineno}

\begin{document}

\title{Large $W$ limit of the knapsack problem}

\author{Mobolaji Williams}
\affil{School of Engineering and Applied Sciences, \\Harvard University, Cambridge, MA
02138, USA\footnote{\href{williams.mobolaji@gmail.com}{williams.mobolaji@gmail.com}}}
\affil{Jellyfish, Boston, MA 02111, USA\footnote{\href{mwilliams@jellyfish.co}{mwilliams@jellyfish.co} }}
\date{March 31, 2023}

\maketitle

\begin{abstract}
We formulate the knapsack problem (KP) as a statistical physics system and compute the corresponding partition function as an integral in the complex plane. The introduced formalism allows us to derive three statistical-physics-based algorithms for the KP: one based on the recursive definition of the exact partition function; another based on the large weight limit of that partition function; and a final one based on the zero-temperature limit of the second. Comparing the performances of the algorithms, we find that they do not consistently outperform (in terms of runtime and accuracy) dynamic programming, annealing, or standard greedy algorithms. However, the exact partition function is shown to reproduce the dynamic programming solution to the KP, and the zero-temperature algorithm is shown to produce a greedy solution. Therefore, although dynamic programming and greedy solutions to the KP are conceptually distinct, the statistical physics formalism introduced in this work reveals that the large weight-constraint limit of the former leads to the latter. We conclude by discussing how to extend this formalism in order to obtain more accurate versions of the introduced algorithms and to other similar combinatorial optimization problems.


\noindent 
\end{abstract}

{%
	\hypersetup{linkcolor=black}
	\setcounter{tocdepth}{2}
    \newpage
	\tableofcontents
}

\newpage
\section{Introduction \label{sec:intro}}


The Knapsack Problem (KP) is a classic problem in combinatorial optimization. In the 0-1 version of the problem \cite{kellerer2004general}, we begin with $N$ objects labeled $i=1, 2, \ldots, N$ where each object can be included or excluded from a collection. When in the collection, the object $i$ has a value $v_i$ and a weight $w_i$, and the objective is to find the combination of objects that maximizes the total value while remaining under a given total weight $W$, called the "weight-limit." A particular collection of objects is defined as $\tbf{x} = (x_1, x_2, \ldots, x_N)$ with $x_i =1$ or $x_i =0$ for object $i$ being included or excluded, respectively, in the collection, and the weight and value vectors are $\tbf{w} \equiv (w_1, w_2, \ldots, w_N)$ and $\tbf{v} \equiv (v_1, v_2, \ldots, v_N)$, respectively. Then the objective in solving the KP is to find $\tbf{x}$ that
\begin{align}
 \text{maximizes $\,\tbf{v} \cdot \tbf{x}\,$ subject to the constraint $\,\tbf{w} \cdot \tbf{x} \leq W\,$},
  \label{eq:knap_constr}
\end{align}
where $\tbf{a} \cdot \tbf{b} \equiv a_1 b_1+ a_2 b_2 + \cdots + a_N b_N$. For simplicity, we take $v_i$, $w_i$, and $W$ to be positive integers. 

Among the many standard algorithms for solving the KP \cite{toth1990knapsack,rosettacodeknapsack2019} there exists a stochastic approach motivated by the physical process of annealing. 
In the simulated annealing approach to the KP, the negative of the total value of the collection of objects is taken to be the energy of the system, and the system evolves by stochastically sampling the possible objects that can be included in the collection consistent with the weight limit. During this evolution, the temperature parameter is slowly (e.g., logarithmically in time \cite{ingber1993simulated}) lowered until the system has settled into its maximum-value collection of objects that is consistent with the weight constraint. 


In this work, we analytically model the physical system at the heart of the simulated annealing solution to the KP. Specifically, we compute the thermal partition function for the KP and derive expressions for the average occupancy of each object. By using contour integral identities, we represent an intuitive summation over a discrete state-space as an abstract integration over a continuous contour in the complex plane, and then, by taking the large $W$ limit of the partition function, we transform our discrete-space optimization objective with a constraint into a continuous-space optimization objective with no constraint. The transformation allows us to optimize our problem using analysis thereby simplifying an originally discrete-space analysis.  A similarly inspired approach was taken in \cite{anderson1988neural} by using a mean field approximation and the saddle-point approximation to model the annealing of a spin-system, and here we use similar methods to derive new algorithms for the KP and then use the formalism to understand the relationships between more well-known algorithms.  

\begin{figure}[t]
\centering
\includegraphics[width=\linewidth]{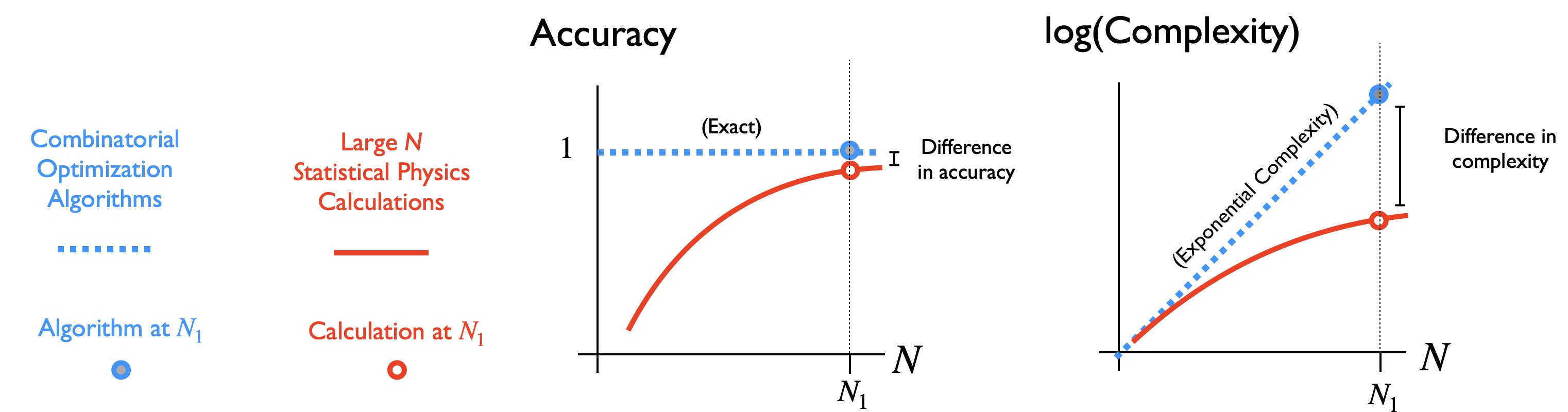}
	\caption{Schematic of accuracy-complexity tradeoff in combinatorial optimization and statistical physics. Combinatorial optimization algorithms are exact, but they often have exponential complexity. $N\gg1$ statistical physics "algorithms" are approximate but become more accurate as $N$ gets larger, and they have sub-exponential complexity since they are based on evaluating expressions rather than on enumerating an exponential number of states. The relatively quick  $N\gg1$ limit physical analog of a combinatorial optimization problem should be most accurate in the same numerical regime where the exact optimization algorithm takes the longest time to compute.}
\label{fig:accuracy_complexity}
\end{figure}

This work exists at the intersection of combinatorial optimization and statistical physics and reflects the theme of gaining insights into the former by representing them as the latter \cite{mezard2009information,percus2006computational}. Importantly, this particular intersection makes use of a convenient confluence between statistical physics and combinatorial optimization: While problems in statistical physics tend to get easier as we increase the number of degrees of freedom $N$ (due to the availability of approximation schemes) problems in combinatorial optimization tend to get harder as we increase $N$ (due to increased computation time (Fig. \ref{fig:accuracy_complexity})). 

The relationship between $N\gg1$ problems in computer science and $N\gg1$ limits in statistical physics has been explored in the past \cite{anderson1986statistical, nishimori2001statistical}, but generally the focus of these works for the KP has been on using annealed approaches or replica methods to study the ground states of physical analogs of combinatorial optimization problems (as in \cite{fontanari1995statistical, korutcheva1994statistical,inoue2007self}) rather than on using statistical physics to find specific solutions to the problems themselves. In the current work, we translate the KP into a physical system with the primary goal of finding a framework for solving the former. 

In Section \ref{sec:part_func}, we implement this translation by first deriving the exact partition function for the system. We show that this partition function can be defined through a recurrence relation that reproduces the standard dynamic programming solution to the KP. In Section \ref{sec:largeN}, we approximate the partition function by applying the method of steepest descent in the large $W$ limit. In Section, \ref{sec:soln} we derive a solution to the KP from the approximated partition function, and this solution in turn leads to two versions of a statistical-physics based algorithm for the KP. In Section \ref{sec:speed}, we compare the performances (in terms of runtime and accuracy) of the introduced algorithms to the performances of other standard solutions to the KP for so-called "difficult instances" of the KP \cite{pisinger2005hard}. We note that the introduced algorithms do not perform better than standard KP algorithms, but in Section \ref{sec:greedy_dynamic} we show that the statistical physics formalism of the new algorithms makes clear relationships amongst the standard ones. In particular, we show that the large $W$ limit of the dynamic programming solution to the KP yields a greedy algorithm in much the same way that the large $N$ limit of the factorial of a number yields Stirling's approximation of said number. We conclude by discussing how higher order corrections could be computed for the introduced algorithms and how the demonstrated connection between dynamic programming and greedy algorithms could be extended to other combinatorial optimization problems.


\section{Partition Function for the Knapsack Problem \label{sec:part_func}}


The partition function is a foundational theoretical construct in statistical physics \cite{mcquarrie73}, and if one can calculate it for a system, then all observables for the system can be calculated in turn. Calculating useful forms of the partition function is difficult for all but the simplest systems \cite{baxter2016exactly}, but there are often approaches to approximating an answer. For the 0-1 KP considered in the body of the paper, we will derive an expression for the partition function and then use the expression to compute the solution to the KP. Making the calculated results computationally useful will require approximations discussed in the next section, but in this section the results will be exact. 

We start by representing the KP as a statistical-physics system at a dimensionless temperature $T$. This temperature is a non-physical hyperparameter that ultimately be taken to zero to ensure that an optimal solution is found. For the KP, this optimal solution corresponds to the highest-value subset of objects that is consistent with the constraint. 

To write the partition function for the KP, we need to place the objective function and the constraint in \rfw{knap_constr} in a sum over all possible states $\tbf{x}$. The possible states consist of all $2^N$ possible vectors $\tbf{x} = (x_1, x_2, \ldots, x_N)$  where $x_j \in \{0, 1\}$ and for which only some vectors satisfy the weight constraint. To impose the constraint, we introduce the Heaviside step function $\Theta(j)$ defined as 
\begin{equation}
\Theta(j) \equiv \begin{dcases} 1 & \text{for $j\geq0$} \\ 0 & \text{otherwise} \end{dcases},
\label{eq:heaviside_def}
\end{equation}
where $j$ is an integer. With \rfw{heaviside_def} and taking the negative of the total value $\tbf{v}\cdot \tbf{x}$ to be the energy of the system, the partition function for the KP is then
\begin{equation}
Z_{N}(\beta \tbf{v}, \tbf{w}, W) = \sum_{\tbf{x}} \Theta\Big( W- \tbf{w}\cdot \tbf{x} \Big) \exp\Big(\beta \tbf{v}\cdot \tbf{x} \Big),
\label{eq:knapsack_pf}
\end{equation}
where $\beta \equiv 1/T$, and the summation over $\tbf{x}$ is defined as 
\begin{equation}
\sum_{\tbf{x}} \equiv \prod_{j=1}^N \sum_{x_j =0}^1 \qquad \text{[Summation for $0$-$1$ problem]}.
\label{eq:01summ}
\end{equation}
On the left hand side of \rfw{knapsack_pf}, the subscript $N$ represents the total number of items under consideration for the KP. With the partition function \rfw{knapsack_pf}, standard statistical physics formalism tells us that the probability that a particular collection of objects $\overline{\tbf{x}}$ occurs in the system is
\begin{equation}
P_N(\bar{\tbf{x}}) = \frac{1}{Z_N} \Theta\Big( W- \tbf{w}\cdot \overline{\tbf{x}} \Big) \exp\Big(\beta \tbf{v}\cdot \overline{\tbf{x}}\Big).
\label{eq:prob_xi}
\end{equation}
Using this probability to express the average occupancy for object $k$ yields
\begin{equation}
\langle x_k \rangle =  \sum_{\tbf{x}} x_k P_N(\tbf{x}) = \frac{\partial}{\partial (\beta v_k)} \ln Z_{N} (\beta\tbf{v}, \tbf{w}, W).
\label{eq:avg_xk} 
\end{equation}
As an average occupancy for this 0-1 problem, \rfw{avg_xk} also represents the probability that object $k$ is included in the collection. For example, if $\langle x_k \rangle > 1/2$, then more than half of the Boltzmann-weighted microstates in the system have object $k$ in the knapsack at a temperature $T$. Therefore, sampling these microstates would give us a greater than $50\%$ chance of having object $k$ included in the collection.


We can go from an "average occupancy" to an explicit prediction of occupancy by taking averages to the zero temperature limit. For a physical system with a set of microstates $\{{\cal S}\}$ and an energy function $-{\cal O}({\cal S})$, defined in terms of the positive-definite objective function ${\cal O}({\cal S})$, the system partition function (at inverse temperature $\beta$) is $ Z = \sum_{\{ {\cal S}\}} \exp\left[\beta {\cal O}(\cal S) \right],$
and the average microstate at this temperature is is 
$\langle {\cal S} \rangle \equiv Z^{-1} \sum_{\{ {\cal S}\}} {\cal S} \exp\left[\beta {\cal O}(\cal S) \right].$
If there is a single microstate ${\cal S}_0$ that maximizes ${\cal O}(S)$, then it is easy to show that
\begin{equation}
\lim_{\beta \to \infty} \langle {\cal S} \rangle = {\cal S}_{0}.
\label{eq:avg_opt}
\end{equation}

Interpreting \rfw{avg_opt} for the KP, the exact solution to the KP is found when \rfw{avg_xk} is taken to the $T \to 0$ (or equivalently when $\beta \to \infty$) limit because as temperature goes to zero, the microstate that dominates the summation is that which maximizes $\tbf{v} \cdot \tbf{x}$ consistent with the constraint $W \geq \tbf{w}\cdot \tbf{x}$. Explicitly, the solution vector $\tbf{X}$ has components defined as 
\begin{equation}
X_{k}^{\text{soln}}  = \lim_{\beta\to \infty} \langle x_k \rangle  = \lim_{\beta\to \infty}\frac{\partial}{\partial (\beta v_k)} \ln Z_{N} (\beta\tbf{v}, \tbf{w}, W).
\label{eq:Xsolndef}
\end{equation}

In the quest to establish a statistical physics based KP algorithm, our local goal is to find an expression for the average occupancy \rfw{avg_xk} and to use this expression within the zero-temperature solution--or low temperature approximations of the solution--\rfw{Xsolndef}. But to move forward, we first need to express the partition function \rfw{knapsack_pf} in a more mathematically useful form. We do so by moving from a discrete-space summation to a continuous-space integration. 

First, we write the Heaviside step function \rfw{heaviside_def} in terms of a contour integral by using the contour integral expression of the kronecker delta $\delta(j, m)$:
\begin{align}
\Theta(j) & = \sum_{m=0}^{\infty}\delta(j, m)  = 
\frac{1}{2\pi i} \oint_{\Gamma} \frac{dz}{z^{j+1} }\sum_{m=0}^{\infty} z^m 
=\frac{1}{2\pi i} \oint_{\Gamma} \frac{dz}{z^{j+1}} \frac{1}{1-z} ,
\label{eq:heaviside_intg}
\end{align}
where, in the final equality, we used the geometric series identity $\sum_{n=0}^{\infty} z^n = {1}/(1-z)$. The identity is only valid for $|z| < 1$, thus applying it constrains the contour $\Gamma$ to not extend more than $1$ unit away from the origin. 
Now, inserting \rfw{heaviside_intg} into \rfw{knapsack_pf}, taking $j \equiv W - \tbf{w}\cdot\tbf{x}$, and summing over the states $\tbf{x}$, we obtain 
\begin{align}
Z_{N}(\beta \tbf{v}, \tbf{w}, W) 
& = \frac{1}{2\pi i} \oint_{\Gamma} \frac{dz}{z^{W+1}} \frac{1}{1-z} \prod_{k=1}^N\Big(1+ z^{ w_k}e^{\beta v_k}\Big).
\label{eq:knapsack_prelim}
\end{align}
\rfw{knapsack_prelim} is the final form of the KP partition function, and it ultimately allows us to solve the associated problem. To obtain this solution, we make explicit one property that will be useful in subsequent discussions. If we isolate the $k=N$ factor in the integrand of \rfw{knapsack_prelim} and expand this factor as two terms, we obtain the identity
\begin{equation}
Z_N(\beta \tbf{v}, \tbf{w}, W)  = Z_{N-1}^{(N)}(\beta \tbf{v}, \tbf{w}, W) + e^{\beta v_N}Z^{(N)}_{N-1}(\beta \tbf{v}, \tbf{w}, W-w_N),
\label{eq:dynamic_prog}
\end{equation}
where $Z^{(k)}_{N-1}(\beta \tbf{v}, \tbf{w}, W)$ is the partition function in which the $\ell$th component is eliminated from both $\tbf{v}$ and $\tbf{w}$, and thus only $N-1$ items are under consideration. Although \rfw{dynamic_prog} was computed for $k=N$, the equality applies for any $k$. In statistical physics, computing the partition function amounts to "solving" the corresponding system, so \rfw{dynamic_prog} shows that solutions to the KP can be built up recursively in terms of the solutions to instances with smaller weight limits and fewer items.  We can write the explicit solution to the KP in terms of this statistical physics representation by applying \rfw{Xsolndef} to \rfw{knapsack_prelim} and using the identity \rfw{dynamic_prog}. We then find 
\begin{equation}
    X_k  = 1 - \lim_{\beta\to\infty}\frac{Z_{N-1}^{(k)}(\beta\tbf{v}, \tbf{w}, W)}{Z_{N} (\beta\tbf{v}, \tbf{w}, W)},
    \label{eq:XsolnSimp}
    \end{equation}
where $Z^{(k)}_{N-1}(\beta \tbf{v}, \tbf{w}, W)$ is the partition function for which the $k$th component is eliminated from both $\tbf{v}$ and $\tbf{w}$. \rfw{XsolnSimp} shows that the explicit solution to the KP can be written in terms of the limit of a ratio of partition functions, and thus being able to compute the partition function \rfw{knapsack_prelim} indeed leads to a solution to the KP.

\begin{algorithm}[t]
    \caption{Exact $Z$ ($T\neq 0$) algorithm}\label{algo:exactz}
    \begin{algorithmic}[1]
        \STATE  Define $\tbf{w} = (w_1, w_2, \ldots, w_N)$, $\tbf{v} = (v_1, v_2, \ldots, v_N)$, and $W$. Select a temperature $T$ satisfying $T\ll \min\{v_j\}$.
        \STATE Define an $(N+1)\times (W+1)$ matrix $Z$ with elements set by $Z[i][w] := 1$ for $0 \leq i \leq N$ and $0\leq w \leq W$. 
        \STATE \tbf{Computing Partition Function:} Compute $Z[N][W]$ as follows: \\[.25em]
        \STATE for $i$ in $[1, N]:$ \\[.25em]
        \STATE $\qquad$ for $w$ in $[1, W]:$\\[.25em]
        \STATE $\qquad$  $\qquad$ if $w_i>w$:\\[.25em]
        \STATE $\qquad$  $\qquad$ $\qquad$  $Z[i][w] := Z[i-1][w]$\\[.5em]
        \STATE $\qquad$ $\qquad$ else:\\[.25em]
        \STATE $\qquad$ $\qquad$ $\qquad$ $Z[i][w] := Z[i-1][w] + \exp(v_i/T)Z[i-1, w-w_i]$\\[.25em]
        \STATE end
        \STATE \tbf{Computuing Solution:} Set $w=W$. Let $\tbf{X}$ represent the final object occupancy for the KP.  Define $\tbf{X}= (X_1, X_2, \ldots, X_N)$ where $X_j:=0$ initially. The values $X_j$ are then updated as follows:\\
        \STATE  for $j$ in $[N, 1]$ (reverse-order list): \\[.25em]
        \STATE  $\qquad$ if $1-Z[j-1][w]/Z[j][w]>1/2$: \\[.25em]
        \STATE  $\qquad$ $\qquad$ $X_j :=1$ \\[.25em]
        \STATE  $\qquad$ $\qquad$ $w :=w-w_j$\\[.25em]
        \STATE end
    \end{algorithmic}
\end{algorithm}

In Algorithm \ref{algo:exactz}, we express the solution \rfw{XsolnSimp} and the partition function recursion \rfw{dynamic_prog} as an algorithm. We call Algorithm \ref{algo:exactz} an "Exact $Z$" algorithm since it is based on computing an exact partition function, but the algorithm itself is an approximate solution to the KP. From the form of \rfw{dynamic_prog} we see that $T$ must be non-zero in order for the partition function to be finite. However, because the exact solution \rfw{XsolnSimp} employs a $T\to0$ limit, any partition-function-based solution to the KP that uses non-zero $T$ will necessarily be an approximate solution. 

Still, we \textit{can} use this formalism to find an exact solution to the KP.  We do so by first highlighting a noteworthy aspect of Algorithm \ref{algo:exactz}: The way the partition function matrix $Z[N][W]$ is built up in line 9 is reminscent of the more traditional dynamic programming solution to the KP \cite{kellerer2004dynamic} in which optimal values for the desired instance are built up from optimal values for subset instances. 

To make the connection between the exact partition function \rfw{knapsack_prelim} and the standard dynamic programming KP solution explicit, we first compute the average total value of the knapsack, $\langle \tbf{v} \cdot \tbf{x} \rangle_{N, W}$, with $N$ items and a weight limit $W$:
\begin{equation}
    \langle \tbf{v} \cdot \tbf{x} \rangle_{N, W} = \frac{1}{Z_N(\beta \tbf{v}, \tbf{w}, W)}\frac{\partial}{\partial \beta} Z_N(\beta \tbf{v}, \tbf{w}, W).
    \label{eq:VNW0}
    \end{equation}
\rfw{VNW0} is a temperature-dependent quantity, but we know (from \rfw{avg_opt}) that taking the zero-temperature limit of such a quantity yields the optimal value of the quantity across the available states. That is, if $V_{N}(W)$ is the optimal value for our instance of the KP, then we must have 
\begin{equation}
    V_{N}(W) = \lim_{\beta\to\infty} \langle \tbf{v} \cdot \tbf{x} \rangle_{N, W}.
    \label{eq:VNW_lim0}
\end{equation}
Now, using \rfw{VNW0} and \rfw{dynamic_prog} in \rfw{VNW_lim0}, we find that the recursive relation \rfw{dynamic_prog} leads (see Appendix \ref{app:dynamic}) to an analogous recursive relation for $V_{N}(W)$:
\begin{equation}
    V_{N}(W) = \begin{dcases} V_{N-1}(W) & \text{for $W< w_N$} \\ \max\{V_{N-1}(W),\, v_{N} + V_{N-1}(W-w_N)\} & \text{for $W\geq  w_N$},\end{dcases}
    \label{eq:bellman}
    \end{equation}
\rfw{bellman} is the standard dynamic programming solution to the KP \cite{kellerer2004dynamic}, and so we have found that the definition of the KP partition function in \rfw{knapsack_pf} implies the validity of the recursion relation \rfw{dynamic_prog} which itself implies that the optimal value of the knapsack (i.e., the average value at zero temperature) is defined recursively by the standard dynamic programming solution. In essence, the exact KP partition function encodes the dynamic programming solution to the KP. 

From this relationship, we see that \rfw{dynamic_prog} and \rfw{bellman} provide us with two equivalent ways of formulating solutions to the KP. We can either recursively compute the paritition function (and then apply \rfw{XsolnSimp}), or we can recursively compute the optimal value. Both approaches amount to the dynamic programming solution to the KP. 
\begin{equation}
\begin{array}{ccc}
\text{Standard Dynamic Programming} &  & \text{Exact Partition Function} \\
 &\cong&  \\
(\textit{Implement \rfw{bellman} recursively to compute $V_{N}(W)$}) &  & (\textit{Implement \rfw{dynamic_prog} recursively to compute $Z_{N}$}) \\
\end{array}
\end{equation}
We will revisit this relationship in Sec. \ref{sec:greedy_dynamic} when we consider an analogous relationship for the greedy algorithm of the KP.

\section{Approximating the Partition Function \label{sec:largeN}}

We have seen that we can use \rfw{dynamic_prog} to implement a dynamic programming solution to compute the partition function for a given $N$ and $W$, and then use  \rfw{XsolnSimp} to find the occupancy of object $j$. However, such a computation is typically slower than the standard DP approach due to the computational complexity of the exponential in \rfw{dynamic_prog}, especially for large arguments. Another approach could involve computing $Z_{N} (\beta\tbf{v}, \tbf{w}, W)$ directly from \rfw{knapsack_pf}, however, this computation would amount to a brute force solution which requires a summation over all $2^N$ states of the system. Alternatively, we could try to use \rfw{knapsack_prelim} to compute the partition function, but this calculation too would be stymied, this time  by the general numerical intractability of the contour integral. 

These challenges suggest we explore an alternative form for the KP partition function in order to more fully capture the algorithmic potential of the statistical physics formulation. This alternative form we find will be based on an approximation of the contour integral. Towards finding this approximation, we first write \rfw{knapsack_prelim} as 
\begin{equation}
Z_{N}(\beta \tbf{v}, \tbf{w}, W)  = \frac{1}{2\pi i} \oint_{\Gamma} \frac{dz}{z} \,\exp F_{N}(z; \beta \tbf{v}, \tbf{w}, W)
\label{eq:knapsack_pf2}
\end{equation}
where we defined 
\begin{equation}
F_{N}(z; \beta \tbf{v}, \tbf{w}, W) = - W \ln z - \ln (1-z) + \sum_{k=1}^N\ln \left(1+ z^{ w_k}e^{\beta v_k}\right).
\label{eq:knapsack_Fdef}
\end{equation}
It is important to note that \rfw{knapsack_Fdef} is not defined for all $z \in \mathbb{C}$; given the condition that defines \rfw{heaviside_intg}, \rfw{knapsack_Fdef} is also only valid for $|z| <1$. 

To approximate \rfw{knapsack_pf2}, we use the method of steepest descent \cite{hassani2013saddlepoint}. The method states that for a given function $f_N: \,\mathbb{R} \to \mathbb{R}$ that obeys $\lim_{N\to \infty} f_N(x) = \infty$, we have 
 \begin{equation}
 I_N= \int_{C} dz\, g(z) e^{f_N(z)} = e^{i\theta_1} \sqrt{\frac{2\pi}{|f''_N(z_0)|}}\, g(z_0) \,e^{f_N(z_0)}(1 + \epsilon_N), 
 \label{eq:saddle_expr}
 \end{equation}
 where $g: \,\mathbb{R} \to \mathbb{R}$; $\epsilon_N$ is an error term; $C$ is a contour in the complex plane; $z_0$ is defined by $f'_N(z_0) = 0$; and $\theta_1$ is defined by the constraint $2\theta_1 + \theta_2 = \pi$ with $\theta_2$ itself defined as the phase of $f''_N(z_0)$ (specifically through $\frac{1}{2} f''_N(z_0) = r e^{i\theta_2}$ for $r, \theta_2 \in \mathbb{R}$; see \cite{hassani2013saddlepoint} for the sources of these phases). 

 In order to apply \rfw{saddle_expr} to \rfw{knapsack_pf2},  we need to identify $g(z)$ and $f(z)$. There is some ambiguity in how we make these identifications, but in general for these steepest descent approximations, one subsumes all factors into the exponential argument aside from the $1/z$ factor typical of contour integrals. By this convention, we can identify $1/z$ with $g(z)$ and $F_N(z; \beta \tbf{v}, \tbf{w}, W)$ with $f_N(z)$\footnote{As an alternative, we could include the $1/z$ factor in our definition of $f_N(z)$ (and also take $g(z) =1$) but then the contribution of the $1/z$ factor to determining the critical point $z_0$ would disappear when we take $W\gg1$ since $(W+1)\ln z \simeq W\ln z$ in this limit.}.

Since $|z|< 1$, we find that $F_N(z; \beta \tbf{v}, \tbf{w}, W)$ goes to $\infty$ when $W\to \infty$. Thus we can infer that $W$ is a suitable large-number parameter, and we can apply the steepest descent approximation \cite{hassani2013saddlepoint} to \rfw{knapsack_pf2} for the case of $W\gg w_j$ for all $j$. Note that taking $W\gg 1$ with no constraint on $w_j$ is not sufficient for ensuring that we are in a regime where the steepest descent approximation applies. An instance satisfying $W\gg 1$ could be trivially created by taking $W \to W' = \lambda W$ and $w_j \to w'_j = \lambda w_j$ for $\lambda\gg1$, but such an instance would be identical to the original instance defined by $W$ and $w_j$.


Applying the steepest descent approximation to \rfw{knapsack_pf2} for $W \gg w_j$, we obtain
\begin{align}
Z_{N}(\beta \tbf{v}, \tbf{w}, W) &  = 
\frac{1}{\sqrt{2\pi z_0^2 \partial^2_z F_N(z_0)}} \exp F_{N}(z_0; \beta \tbf{v}, \tbf{w}, W)\left(1 + \mathcal{O}(w_j/W)\right) ,
\label{eq:knapsack_pf_approx}
\end{align}
where $z_0$ is the value of $z$ at which $\partial_{z}F_N(z; \beta \tbf{v}, \tbf{w}, W) =0$.  Note that to obtain \rfw{knapsack_pf_approx}, we assumed $\partial_{z}^2F_{N}(z_0)$ to be real and positive, thus giving us $\theta_2 = 0$ which in turn implies $\theta_1 = \pi/2$ and $e^{i\theta_1}/i = 1$. Below we verify this assumption. The error term ${\cal O}(w_j/W)$ comes from the fact that the error associated with the approximation is of the order of the inverse of the large-number parameter (see equation 11.33 in  \cite{hassani2013saddlepoint} for a full expansion). 

In what follows, we will take $F_N(z) \equiv F_{N}(z; \beta \tbf{v}, \tbf{w}, W)$ where notationally convenient. Using \rfw{knapsack_Fdef}, we find that the $z$ derivative of $F_N$ is 
\begin{equation}
\partial_{z}F_{N}(z; \beta \tbf{v}, \tbf{w}, W) = -\frac{W}{z}+\frac{1}{1-z} +\frac{1}{z}\sum_{i=1}^{N}\frac{w_iz^{w_i}e^{\beta v_i}}{1 +z^{w_i}e^{\beta v_i}}.
\label{eq:Fpartial}
\end{equation}
Therefore, the condition that defines $z_0$ in \rfw{knapsack_pf_approx} is 
\begin{equation}
W = \frac{z_0}{1-z_0} + \sum_{i=1}^{N}\frac{w_iz_0^{w_i}e^{\beta v_i}}{1 +z_0^{w_i}e^{\beta v_i}
}.
\label{eq:Wsoln_z0}
\end{equation}
We can ensure that such a $z_0$ always exists as follows. Taking the limits of \rfw{Fpartial} at the bounds of the domain $z \in (0, 1)$, we have $\lim_{z\to 0} \partial_{z}F_{N}(z) = -\infty$, and $\lim_{z\to 1} \partial_z F_{N}(z) = +\infty$ which implies that $\partial_z F_N(z)$ crosses the axis at some point between $z=0$ and $z=1$. Therefore, by the intermediate value theorem \cite{rudin1964principles} there must exist some $z_0 \in (0, 1)$ such that $\partial_z F_N(z_0) = 0$. 

Next we check that $\partial_{z}^2F_{N}(z_0)$ is real and positive. Differentiating \rfw{Fpartial} and setting $z= z_0$ we find
\begin{align}
\partial_{z}^2F_{N}(z_0; \beta \tbf{v}, \tbf{w}, W) 
 & = \frac{1}{z_0(1-z_0)^2}+  \frac{1}{z_0^2} \sum_{i=1}^{N} \frac{w_i^2z_0^{w_i} e^{\beta v_i}}{(1+ z_0^{w_i}e^{\beta v_i})^2}
 \label{eq:second_partial}
\end{align}
which, for $z_0 \in (0, 1)$, is a positive real number. Thus \rfw{knapsack_pf_approx} is a real quantity, and the $z_0$ determined from \rfw{Wsoln_z0} defines a local minimum for $F_N(z)$. 

Finally, we show that $z_0$ is unique. First assume the contrary, namely that there are two distinct critical points $z_A$ and $z_B$ that satisfy the critical point condition, i.e., $\partial_z F_N(z_A) = \partial_z F_N(z_B)=0$ and $z_A \neq z_B$. By \rfw{second_partial}, we deduce that both of these critical points are local minima. However, if a single-variable function has two local minima at distinct points it must also have a local maximum between those points. Thus, if there are two critical points, there must be a third critical point $z_C$ defining a local maximum. However, we showed in \rfw{second_partial} that if $z_C$ is a critical point of $F_N(z)$, then $z_C$ must be a local minimum. We have found that in order for $z_C$ to exist it must be both a local minimum and a local maximum which is impossible for a single-variable function. Therefore, the initial assumption that there can be two distinct critical points is false, and the critical point $z_0$ must be unique. 

Given that $z_0$ exists and is unique, we can use \rfw{knapsack_pf_approx} to find a unique approximation to the KP partition function. Such an approximation is useful because it allows us to study the statistical physics of the KP without the summation over the $2^N$ microstates that defines the initial form of the partition function \rfw{knapsack_pf}. In particular, with \rfw{avg_xk} and \rfw{Xsolndef} we know that the solution to the KP can be expressed in terms of derivatives of this partition function, and consequently, an approximate solution to the KP can be expressed in terms of derivatives of the approximate partition function \rfw{knapsack_pf_approx}. We pursue this approximate solution in the next section. 

\section{{Solving the Knapsack Problem} \label{sec:soln}} 

Having obtained the unique $z_0$ determined by the condition \rfw{Wsoln_z0}, we can use \rfw{avg_xk} and \rfw{knapsack_pf_approx} to find an approximate expression for $\langle x_\ell \rangle$ and in turn use this expression to solve the KP. We first consider nonzero-temperature solutions to the KP, and then show how these solutions lead to a zero-temperature approach. 

\subsection{Nonzero-Temperature \label{sec:nonzeroT}}

We seek to use the large $W$ approximation results  \rfw{knapsack_pf_approx} to explicitly solve the KP. We begin by writing $\langle x_{\ell} \rangle$ in terms of $F_N(z)$ as 
\begin{align}
\langle x_{\ell} \rangle = \frac{\partial}{\partial (\beta v_{\ell})} \ln Z_N(\beta \tbf{v}, \tbf{w}, W) = \frac{\partial}{\partial (\beta v_{\ell})}F_{N}(z; \beta \tbf{v}, \tbf{w}, W)+{\cal O}(w_j/W), 
\end{align}
where we ignored derivatives with respect to the prefactors in the approximation because they are sub-leading in our large $W$ limit. Calculating this quantity from \rfw{knapsack_Fdef}, we obtain 
\begin{equation}
\langle x_{\ell} \rangle  =
 \frac{ z_0^{w_\ell} e^{\beta v_\ell}}{1+z_0^{w_\ell} e^{\beta v_\ell}}+{\cal O}(w_j/W).
\label{eq:kn_soln}
\end{equation}
Assuming we can find $z_0$ with \rfw{Wsoln_z0},  we can use \rfw{kn_soln} to determine the average occupancy for the object $\ell$. We can then use all of these approximate result as the basis for a large $W$ finite temperature algorithm for the KP. To formulate the algorithm, we assume that we have a black-box solver that can obtain the numerical solution to a nonlinear equation (e.g., \cite{nsolvewolfram,scipyfsolve}). We denote this black-box solver as $\texttt{NSolve}$, and, notationally, we write $x_0 = \texttt{NSolve}(x; F(x))$ when $x_0$ is the solution the algorithm finds to the equation $F(x) = 0$. As established in the Sec. \ref{sec:largeN}, there is only one solution to \rfw{Wsoln_z0}, so we do not need to be concerned with multiple roots. The average in \rfw{kn_soln} yields a value between $0$ and $1$, exclusive, and so in order to convert the result into an unambiguous solution we need to introduce a paramater $p_{\text{thresh}}$ that defines how large $\langle x_{\ell} \rangle$ needs to be in order for object $\ell$ to be included in the collection. In Algorithm \ref{algo:tneq0}, we formulate these ideas as a sequence of steps that yields an approximate solution to the KP.  

\begin{algorithm}[t]
    \caption{Large $W$  nonzero-temperature algorithm}\label{algo:tneq0}
    \begin{algorithmic}[1]
        \STATE  Define $\tbf{w} = (w_1, w_2, \ldots, w_N)$, $\tbf{v} = (v_1, v_2, \ldots, v_N)$, and $W$. Select the system temperature $T$ satisfying $T\ll \min\{v_\ell\}$, and a probability threshold $p_{\text{thresh}}$.
	\STATE Compute $z_0(T)$ by using a numerical solver $\texttt{NSolve}$ to evaluate
\begin{equation}
z_0 = \texttt{NSolve}(z; G_{N}(z;  \tbf{v}/T, \tbf{w}, W) ),
\end{equation}
where 
\begin{equation}
G_N(z;\tbf{v}/T, \tbf{w}, W) = -W + \frac{z}{1-z} + \sum_{i=1}^{N}\frac{w_{i}}{ e^{- v_i/T}z^{-w_i} +1},
\label{eq:GNz}
\end{equation}            
at the chosen $T$.
\STATE Compute $\langle x_{\ell} \rangle$ from 
\begin{equation}
\langle x_{\ell} \rangle  = \frac{1}{ e^{- v_\ell/T}z_0^{-w_\ell} +1}.
\label{eq:xellsoln}
\end{equation}
\STATE For each $\ell = 1, \ldots, N$ compute $X_{\ell}$ according to 
\begin{equation}
X_{\ell} = \begin{dcases} 1 & \text{if $\langle x_{\ell} \rangle> p_{\text{thresh}}$,} \\ 0 & \text{otherwise.}\end{dcases}
\label{eq:Xelldef}
\end{equation}
The vector $\tbf{X} = (X_1, X_2, \ldots, X_N)$ represents the final object composition.
    \end{algorithmic}
\end{algorithm}





\begin{figure*}[t!]
    \centering
    \begin{subfigure}[t]{0.5\textwidth}
        \centering
        \includegraphics[width = \linewidth]{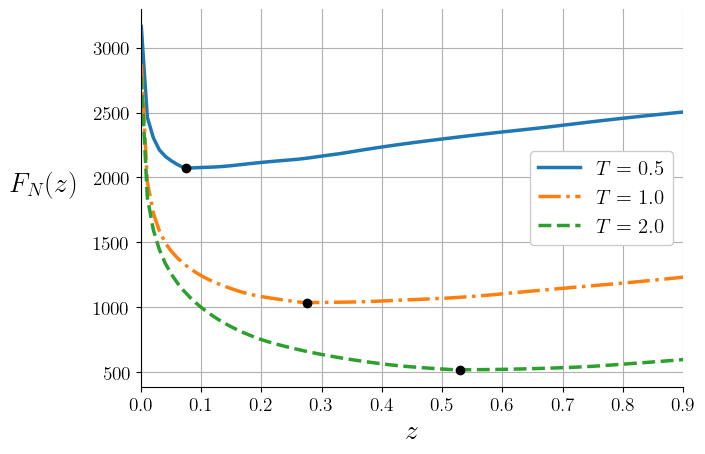}
        \caption{Plot of  \rfw{knapsack_Fdef} at various temperatures}
        \label{fig:roll_down}    
    \end{subfigure}        $\qquad$
    \begin{subfigure}[t]{0.425\textwidth}
        \centering
        \includegraphics[width = \linewidth]{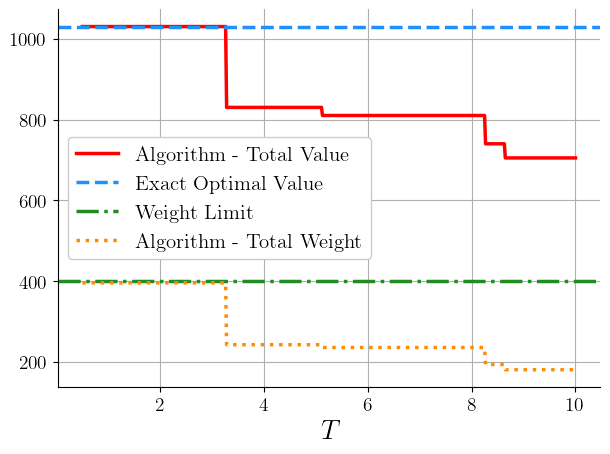}
        \caption{Calculated value and weight as a function of temperature}
        \label{fig:v_opt}    
    \end{subfigure}    
    \hfill
    \caption{Temperature dependence of large $W$ algorithm: Plots correspond to a KP instance with $N=22$ and a $W=400$. Exact weights and values are provided in the code referenced in \textit{Data Availability Statement} along with all the functions used to generate the figures.  (a)  In the $T\neq0$ algorithm, we are "rolling down" the hill represented by $F_N(z)$ (\rfw{knapsack_Fdef}) and into the local minimum (marked as black circle). As the temperature of the system is lowered, $z_0$ decreases.  (b) Plot of $\tbf{v} \cdot \tbf{X}$ and  $\tbf{w} \cdot \tbf{X}$ for the $X_{\ell}$ computed from \rfw{Xelldef} with $p_{\text{thresh}} = 0.95$;  As we lower the temperature, the optimized value increases as does the associated weight, until we reach the weight limit.}
    \label{fig:soln_temp_dependence}
\end{figure*}

In Fig. \ref{fig:v_opt}, we show the results of applying this algorithm to an example instance. As the temperature $T$ of the system is lowered, the total weight of the included objects predicted from \rfw{Xelldef} approaches the limiting weight, and the total value increases to its optimal value. This is as we should expect: In statistical physics, as system temperature $T$ is lowered, the space of microstates where the system spends most of its time gets smaller, until, at zero temperature, the system settles into the single lowest-energy microstate, presuming such a microstate exists. 

It is worth comparing Algorithm \ref{algo:tneq0} with simulated annealing, another algorithm extending from statistical physics. In simulated annealing, a typically discrete state space is explored by randomly proposing and then rejecting or accepting state transitions in an energy landscape with gradually deepening valleys \cite{ingber1993simulated}. The system temperature, which is lowered over time, parameterizes the deepening of the valleys, and as these valleys become more pronounced, the system eventually settles into one of its local minima. In this way, simulated annealing, allows us to computationally find local optima in state spaces. 

These aspects of simulated annealing have analogs in the large $W$ algorithm. In simulated annealing the objective function is the negative of the total value of the object collection, while in Algorithm \ref{algo:tneq0} the objective function is the more abstract complex potential $F_{N}(z)$. In simulated annealing, the algorithm takes small random steps in the direction of the the local minimum of the objective function, while in Algorithm \ref{algo:tneq0}, solving for $z_0$ amounts to directly going to the minimum of the corresponding objective function. The two algorithms have consonant interpretations of how the optimization proceeds, even though each is searching different state spaces with different objective functions. 

Moreover, the "minimmum-seeking" interpretation of Algorithm \ref{algo:tneq0} suggests a connection to a standard approximation in combinatorial optimization. In finding $z_0$ (the minimum of $F_N(z)$), we are effectively "rolling down" the potential defined by $F_N(z)$ and settling at its lowest point (See Fig. \ref{fig:roll_down}). In fact, we could rewrite Algorithm \ref{algo:tneq0} as a gradient descent algorithm to highlight this "valley rolling" interpretation. Such an interpretation reveales that the algorithm is essentially a "greedy approach" in $z$-space for finding the minimum of the potential $F_{N}(z)$. The KP algorithm itself has a greedy solution, so it is worth asking whether there is any relationship between the well-known greedy heuristic for the KP and the greedy optimization of the complex potential $F_N(z)$ in $z$-space. Such a relationship would allow us to explicitly connect the exact dynamic programming solution to the KP (shown in Section \ref{sec:part_func} to be derivable from the partition function) to a greedy solution. We explore this possibility in Sec. \ref{sec:greedy_dynamic}.







\subsection{Zero-Temperature \label{sec:zeroT}}

In Fig. \ref{fig:v_opt}, we saw that the KP solution given by \rfw{xellsoln} became more accurate as system temperature decreased. This behavior suggests a question about the initial solution \rfw{kn_soln}: Can this solution be taken all the way to $T\to0$ in order to obtain the best approximation this formalism can give for the KP? 

To answer this question, we return to the definition of the KP solution in \rfw{Xsolndef}. This definition shows that we can go from a temperature-dependent average occupancy to the solution of the corresponding optimization problem by taking $\beta \to \infty$. Writing \rfw{kn_soln} in terms of the $T\to0$ limit (for notational convenience), we have 
\begin{align}
X^{\text{soln}}_{\ell}  = \lim_{T\to 0} \,\,\frac{1}{e^{-(v_\ell- \gamma(T)w_\ell)/T}+1}+{\cal O}(w_j/W),
\label{eq:Xlim}
\end{align}
where we defined
\begin{equation}
\gamma(T) \equiv -T \ln z_0(T),
\label{eq:gamma_def}
\end{equation}
and we wrote $z_0 = z_0(T)$ to make explicit $z_0$'s dependence on temperature $T$. To use \rfw{Xlim} to obtain a viable solution to the KP, we need to assume that the temperature limit is both non-trivial and well-defined. In other words, we assume that the result is neither zero nor infinite and is instead an explicit function of the arguments $v_{\ell}$, $w_{\ell}$ and an implicit function of the parameters that determine $z_0(T)$. The $1/T$ factor in the argument of the exponential of \rfw{Xlim} suggests a final form for the $T\to0$ limit of \rfw{Xlim}. From the definition of the Heaviside step function $H(x)$ with $H(0) = 1/2$, we have
\begin{equation}
H(x) = \lim_{\alpha \to 0} \frac{1}{e^{-x/\alpha}+1} = \begin{dcases}  1 & \text{for $x>0$}\\ 0 & \text{for $x<0$} \end{dcases}.
\end{equation}
Taking the analogous limit in \rfw{Xlim} then gives us
\begin{equation}
X^{\text{soln}}_{\ell} = H\left(v_\ell/w_{\ell}-  \gamma_0 \right)+{\cal O}(w_j/W), 
\label{eq:hv_soln}
\end{equation}
where we used the identity $H(ax) = H(x)$ for $a>0$ and defined 
\begin{equation}
\gamma_0 \equiv -\lim_{T\to 0}T \ln z_0(T).
\label{eq:gamma_lim}
\end{equation}


The process for implementing the solution represented by \rfw{hv_soln} is simple, in principle: Given $\tbf{v}$, $\tbf{w}$, and $W$, we use the constraint \rfw{Wsoln_z0} to first determine $z_0(T)$, then determine the function $\gamma(T) = - T\ln z_0(T)$, take this function to the $T\to 0$ limit to obtain $\gamma_0$, and finally (by \rfw{hv_soln}), include in our KP solution all objects whose $v_{\ell}/w_{\ell}$ ratio is greater than the value of $\gamma_0$.

As a sanity check for this formalism, we will apply it to the trivial "degenerate" KP instance where all $N$ objects have the same weight $w_0$ and the same value $v_0$. Such an instance does not have a unique solution, and we expect the derived solution \rfw{hv_soln} to reflect this. For the degenerate instance, \rfw{Wsoln_z0} simplifies to 
\begin{equation}
W = \frac{z_0}{1-z_0} + \frac{Nw_0}{e^{-v_0/T} z_0^{-w_0} +1}.
\label{eq:Wtrivial}
\end{equation}
In order to solve the KP with \rfw{hv_soln}, we need to determine $\gamma_0$. The quantity $\gamma_0$ is defined in terms of the zero-temperature limit of $z_0(T)$ and $z_0(T)$ goes to $0$ as $T$ goes to zero. Also, the approximation assumes $W\gg w_j\geq1$. With these two facts, \rfw{Wtrivial} can be approximated by dropping the first term (which is $O(z_0)$) on the right-hand-side. By dropping this term, \rfw{Wtrivial} becomes soluble for this degenerate instance, and we find
\begin{equation}
z_0(T) = \frac{W}{Nw_0-W}e^{-v_0/w_0T} + {\cal O}(z_0^2),
\label{eq:z0T}
\end{equation}
Computing $\gamma(T)$ from \rfw{gamma_def} and \rfw{z0T} together and then taking the limit of the result as $T\to 0$, we find $\gamma_0 = v_0/w_0$, which by \rfw{hv_soln} yields the solution
\begin{equation}
X_{i} = 1/2, \qquad \text{[Degenerate solution]}
.\label{eq:degensoln}
\end{equation}
for all $i$. We expect $X_i$ to be either 1 or 0 to indicate that object $i$ is included or excluded, respectively, from the collection that solves the KP. A value of $1/2$ is therefore ambiguous and such an ambiguity implies that there are multiple viable solutions to the KP for each object: Some solutions where the object is included in the collection and other solutions where the object is excluded from the collection. This structure of solutions is of course true for the degenerate KP since all objects are equivalent and can be switched out of the solution. Therefore the solution \rfw{hv_soln} reproduces what we expect for the degenerate KP\footnote{The value of $X_{\ell}=1/2$ suggests that there is one set of microstate solutions where $\ell$ is included in the collection and \textit{an equal number} of microstate solutions where $\ell$ is excluded from the collection. In fact, the exact multiplicity of solutions is a bit different. With $M \equiv \floor{W/w_0}$, we find there are $\binom{N}{M}$ solutions to the degenerate KP of which $\binom{N-1}{M-1}$ include an arbitrary object $\ell$ and $\binom{N-1}{M}$ do not include that object $\ell$. Therefore, the exact statistical value of $X_{\ell}$ is actually $\binom{N-1}{M-1}/\binom{N}{M} = M/N$. This value is still a fraction, so it is indeed ambiguous as a representation of whether object $\ell$ is included in the collection, but it is different from the value of $1/2$ obtained from the large $W$ result \rfw{hv_soln}. This difference suggests that the value of $X_{\ell} = 1/2$ only implies "solution ambiguity" and does not give the exact microstate proportionality associated with that ambiguity.}.

This degenerate instance was special in that we were able to find an analytical expression for $z_0(T)$ for low temperature. However, for non-trivial instances, determining $\gamma_0$ from the limit \rfw{gamma_lim} is difficult due to the need to find low-temperature solutions to \rfw{Wsoln_z0}, a task which is made difficult due to overflow errors from the exponential function. So instead of determining $\gamma_0$ directly from the $T\to0$ limit of $z_0(T)$, we consider a simpler approach based on the $T\to 0$ limit of the entire expression \rfw{Wsoln_z0}. Using \rfw{hv_soln} and the fact that $\lim_{T\to 0}z_0(T) = 0$, we find that the $T\to 0$ limit of \rfw{Wsoln_z0}) is 

\begin{equation}
    W = \sum_{j=1}^N w_j H(v_j/w_j- \gamma_0).
    \label{eq:Wgreedy}
    \end{equation}
\rfw{Wgreedy} represents a consistency equation for $\gamma_0$, and given $\tbf{v}$, $\tbf{w}$, and $W$, we can solve this equation for $\gamma_0$. \rfw{Wgreedy} is not always soluble in this way, and, in such cases, $\gamma_0$ would be chosen so as to yield the smallest error in the expression. For example, applying \rfw{Wgreedy} to the degenerate instance, yields $W/Nw_0 = H(v_0/w_0 - \gamma_0)$ which yields the solution $\gamma_0 = v_0/w_0$ only if $W/Nw_0 =1/2$. However, regardless of the value of $W/Nw_0$, the constraint of $0<W/Nw_0< 1$ implies that the $\gamma_0=v_0/w_0$ solution is more consistent with this constraint than other values of $\gamma_0$.

\begin{algorithm}[t]
    \caption{Large $W$ zero-temperature algorithm}\label{algo:t=0}
    \begin{algorithmic}[1]
        \STATE  Define $\tbf{w} = (w_1, w_2, \ldots, w_N)$, $\tbf{v} = (v_1, v_2, \ldots, v_N)$, and $W$.
\STATE Approximate $\gamma_0$ by using a numerical solver $\texttt{NSolve}$ to evaluate 
\begin{equation}
    \gamma_0 = \texttt{NSolve}(\gamma; L(\gamma; \tbf{v}, \tbf{w}, W)  ),
    \end{equation}
where
\begin{equation}
    L(\gamma; \tbf{v}, \tbf{w}, W) = W - \sum_{j=1}^N w_j H(v_j/w_j - \gamma ).
    \label{eq:LWconstr}
\end{equation}
\STATE For each $\ell = 1, \ldots, N$ compute $X_{\ell}$ according to 
\begin{equation}
X_\ell = H(v_\ell/w_{\ell}-  \gamma_0),
\label{eq:Xalgsoln}
\end{equation}
where $H(x)$ is the heaviside step function with $H(0) = 1/2$. The vector $\tbf{X} = (X_1, X_2, \ldots, X_N)$ represents the final object composition. For the case of the ambiguous solution $X_{j}=1/2$, we take $X_{j} \to 0$ as an item placement decision. 
    \end{algorithmic}
\end{algorithm}

In Algorithm \ref{algo:t=0}, we formulate a solution to the KP based on \rfw{Wgreedy}. The algorithm is similar to Algorithm \ref{algo:tneq0} except that rather than seeking the value of $z$ at the potential minimum, we are seeking $\gamma_0$ explicitly defined in \rfw{gamma_def} and implicitly defined in \rfw{Wgreedy}. From its use in \rfw{Wgreedy}, the quantity $\gamma_0$ is imbued with a simple conceptual interpretation. With $H$ constrained to be either $1$ or $0$ (i.e., no ambiguous solutions), it is apparent that \rfw{Wgreedy} represents a KP solution in which all items are included in the knapsack if their value-to-weight ratios exceed $\gamma_0$. Thus, $\gamma_0$ is the minimum value-weight ratio that determines knapsack occupancy. 

In the next section, we will take all three introduced algorithms and compare them with some standard algorithms for the KP. The goal in this comparison is to evaluate the claim made in Fig. \ref{fig:accuracy_complexity} and to see whether a statistical-physics based approach to the KP yields solutions with higher accuracy for larger instances while taking comparatively less time than exact algorithms. 

In this work, we have only considered the 0-1 KP, but there are many variations to this classic case. In Appendix \ref{app:var} we consider three variations (the bounded, unbounded and continuous KPs) and show how the statistical physics partition function can be computed for each one. For the bounded and unbounded problems, we also show how this partition function can lead to large $W$ algorithms for the KP.

\section{Runtime and Accuracy Comparisons \label{sec:speed}}

In the introduction, we argued that the large-number limit of the KP should (as the large-number parameter increased) yield solutions that are progressively more accurate while taking relatively less time than an exact algorithm based on brute-search or dynamic programming. In this section, we explore whether this is the case by comparing the introduced algorithms to standard KP algorithms.

Three new algorithms have been introduced in this work: The "exact $Z$ algorithm" (Algorithm \ref{algo:exactz}) based on a calculation of the KP partition function through the recursive definition \rfw{dynamic_prog}; the "large $W$ nonzero-temperature algorithm" (Algorithm \ref{algo:tneq0}) based on the sigmoidal solution for $\langle x_{\ell}\rangle$ in \rfw{kn_soln}; the "large $W$ zero-temperature algorithm" (Algorithm \ref{algo:t=0}) based on the Heaviside step function solution \rfw{hv_soln}. Comparing the runtime and accuracy performances of these algorithms would make evident the relative benefits of the higher accuracy of the exact algorithm versus the faster runtimes of the approximate algorithms in addition to the effect of the $T\to0$ limit on the accuracy of the approximate algorithm. 

Three other standard KP algorithms were used as baseline comparisons. The dynamic programming (DP) algorithm for the KP that has time complexity $O(NW)$; the greedy algorithm which arranges objects as a decreasing sequence in $v_i/w_i$ and then includes objects in the collection according to this sequence until the weight limit is reached \cite{dantzig1957discrete}; the simulated annealing algorithm in which the KP is represented as a thermal system whose temperature is gradually lowered until the system settles into the microstate of highest total value consistent with the constraint \cite{vcerny1985thermodynamical}.

All six algorithms were applied to four different types of "difficult" KP instances taken from \cite{pisinger2005hard}. In \cite{pisinger2005hard} it was noted that the easiest instances of the KP consist of those for which item values are uncorrelated with item weights since such instances very likely contain "obviously-included" items with large values and small weights. A follow-up study \cite{smith2013proteoform} further clarified that difficult KP instances occur when there is a strong correlation between the values and weights or when there are degeneracies in the values and weights of the list of items. In this section we show the results of applying the six algorithms to the "circle" and "spanner" instances, instances which variously exhibix strong correlations or degeneracies. The results for two other difficult instances from \cite{pisinger2005hard}--the "profit-ceiling instance" and the "multiple-strongly correlated items instance"--are discussed in Appendix \ref{app:additional_instances}. 

The circle and spanner instances are constructed as follows. 

\begin{itemize}
\item \tbf{Circle Instances:} Values as a function of the weights form the arc of an ellipse. Taking the weights $w$ to be uniformly distributed in the range $[1,R]$, for a free integer paramater $R$, the values satisfy $v = d \sqrt{4R^2 - (w-2R)^2}$
where $d$ is also a free parameter. In \cite{pisinger2005hard}, it is noted that particularly difficult instances come from choosing $d=2/3$. The value of $R$ was set to $100$.
 



\item \tbf{Spanner Instances:} All items are multiples of a small set of items termed the "spanner set." To create the instance, we first select $\nu$ weights $w_j$ in the interval $[1, R]$, for a free integer parameter $R$, and then define the corresponding values as $v_j = w_j + R/10$. The $N$ items of the instance are found by randomly selecting an item from the $\nu$ items in the spanner set, and randomly selecting an integer $a$ from the interval $[1, m]$, for a free integer paramater $m$. The new item then has the value and weight $(a v_j, aw_j)$, and this is repeated until $N$ items are selected. In what follows, we set $\nu=2$ and $m=10$, in accordance with the construction in \cite{pisinger2005hard}. The value of $R$ was set to $10$. 
\end{itemize}



To compare the set of six algorithms, we considered instances with $N = 2^{k}$ objects for $k=3, 4, \ldots, 12$. For each $N$, integer weights were randomly selected and values were calculated according to the instance definitions given above. The weight limit was set to $\sum_{j=1}^N w_j/2$ which meant that an increase in $N$ led to an increase in $W$, with a mostly linear scaling. Having both $W$ and $N$ increase together was important for maintaining the existence of non-trivial solutions: In order to see the effects of large $W$ on the accuracies of solutions, we were primarily interested in considering instances with increasing $W$,  however if $W$ increased without an increase in $N$, then, given a fixed range of variation for the weights $w_j$, the instance would eventually admit the trivial solution in which all items are included. 

\begin{figure*}[t!]
    \centering
    \begin{subfigure}[t]{0.32\textwidth}
        \centering
        \includegraphics[width = \linewidth]{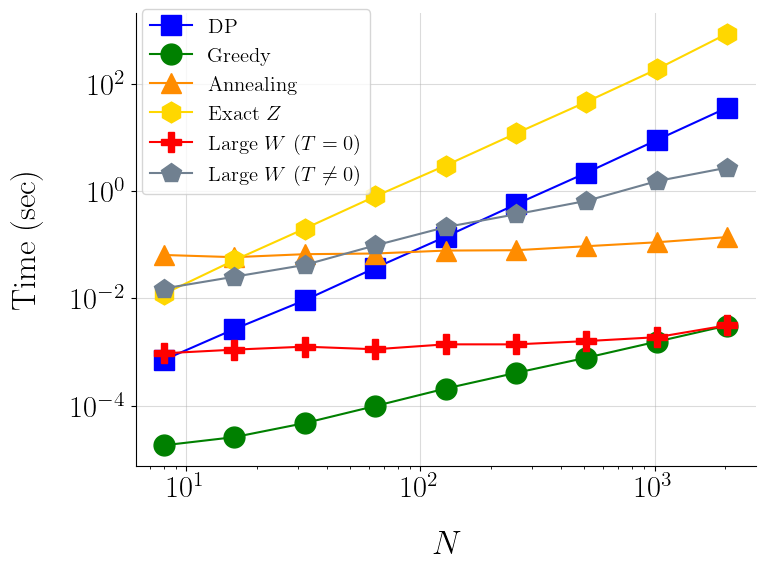}
        \caption{Circle: Time vs. $N$}
        \label{fig:time_n1}
    \end{subfigure} \hfill
    \begin{subfigure}[t]{0.32\textwidth}
        \centering
        \includegraphics[width = \linewidth]{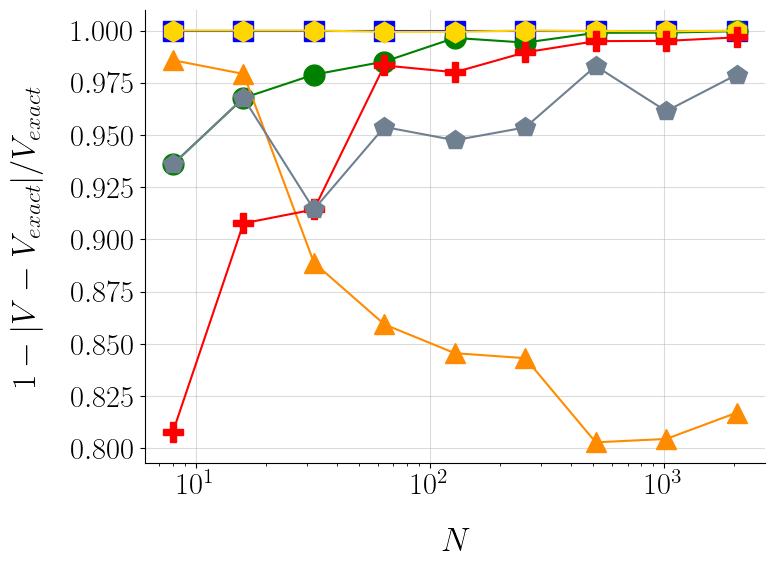}
        \caption{Circle: Acc. vs. $N$}
        \label{fig:v_n1}
    \end{subfigure}\hfill
    \begin{subfigure}[t]{0.32\textwidth}
        \centering
        \includegraphics[width = \linewidth]{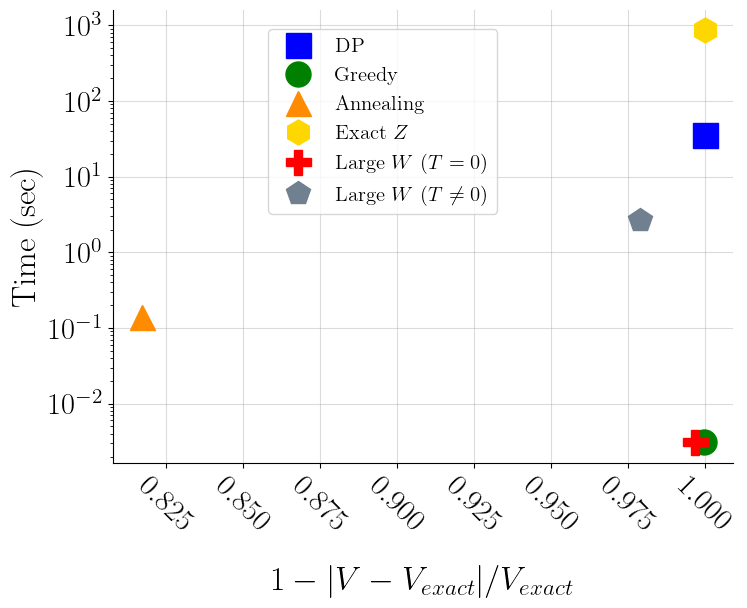}
        \caption{Circle: Time vs. Acc.}
        \label{fig:time_v1}
    \end{subfigure}\\\vspace{.5cm}
    \begin{subfigure}[t]{0.32\textwidth}
        \centering
        \includegraphics[width = \linewidth]{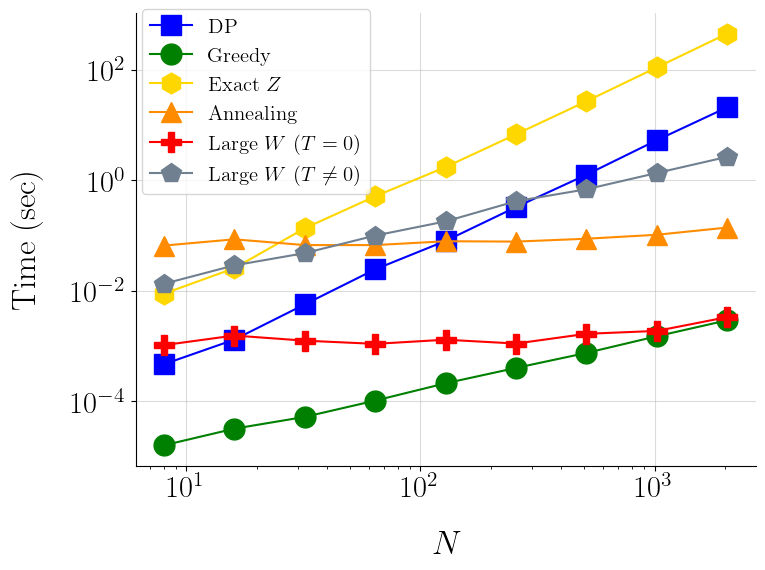}
        \caption{Spanner: Time vs. $N$}
        \label{fig:time_n2}
    \end{subfigure} \hfill
    \begin{subfigure}[t]{0.32\textwidth}
        \centering
        \includegraphics[width = \linewidth]{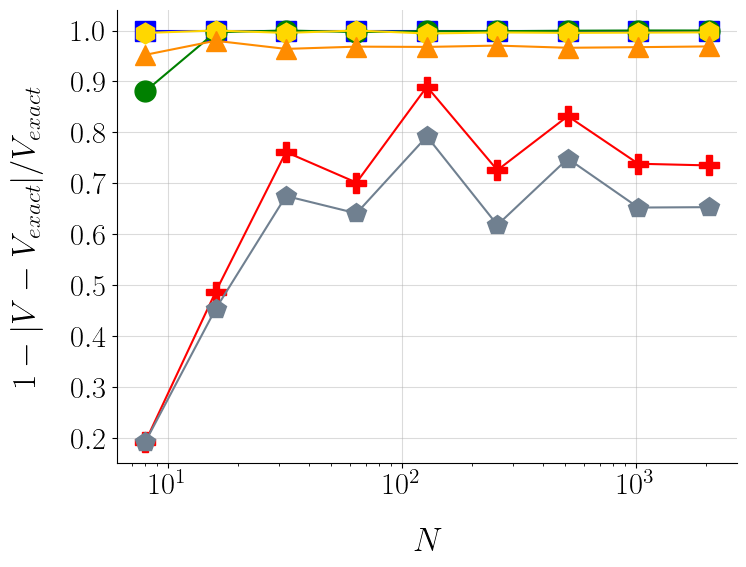}
        \caption{Spanner: Acc. vs $N$}
        \label{fig:v_n2}
    \end{subfigure}\hfill
    \begin{subfigure}[t]{0.32\textwidth}
        \centering
        \includegraphics[width = \linewidth]{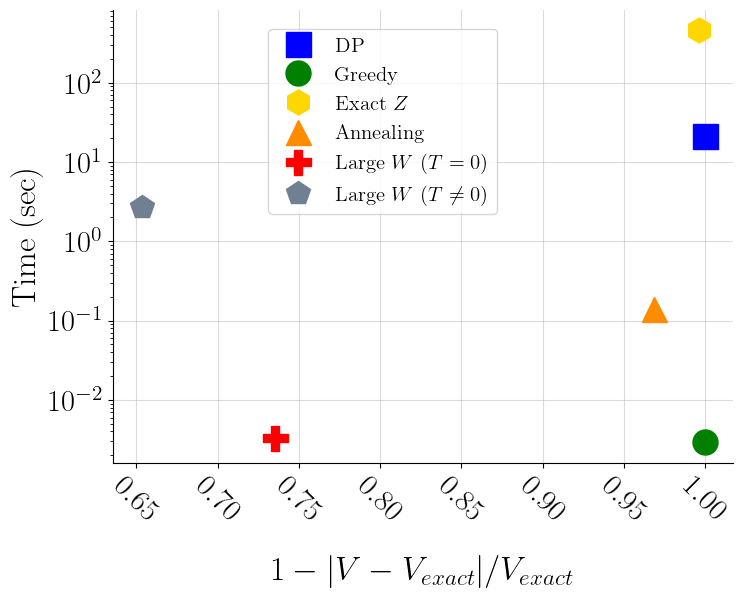}
        \caption{Spanner: Time vs. Acc.}
        \label{fig:time_v2}
    \end{subfigure}    
    \caption{Runtime and Accuracy Comparisons: In (a),(b), and (c), we show the results for the "circle instances" and in (d), (e), and (f), we show the results for the "spanner instances." (a) and (d) depict the runtimes as a function of $N$. (b) and (e) depict the accuracy of the algorithms as a function of $N$. (c) and (f) depict the runtime and accuracy for the instance with $N=2048$ items. For the temperature dependent algorithms we set $T=1.0$. The large $W$ algorithms can achieve high accuracies for the strongly correlated circle instance but fail for the highly degenerate spanner instance. Code used to generate these figures is linked to in Sec. \ref{sec:code}, \textit{Data Availability Statement}.
    }
    \label{fig:spd_accy}
\end{figure*}

The runtime and accuracy results of applying the algorithms to the two difficult instances are presented in Fig. \ref{fig:spd_accy}. Fig. \ref{fig:time_n1}, \ref{fig:v_n1}, and \ref{fig:time_v1} displays the results for the circle instance, and in Fig. \ref{fig:time_n2}, \ref{fig:v_n2}, and \ref{fig:time_v2} displays the results for the spanner instance. Fig. \ref{fig:time_n1} and \ref{fig:time_n2} shows the runtime of each algorithm as function of the total number of items. Fig. \ref{fig:v_n1} and \ref{fig:v_n2} shows the accuracy of each algorithm as a function of the total number of items. Fig. \ref{fig:time_v1} and \ref{fig:time_v2} shows the runtime and accuracy for each algorithm for the case of $N=2048$ items. Since the dynamic programming solution is exact, we defined the accuracy of an algorithm as $1- |V- V_{\text{exact}}|/V_{\text{exact}}$ where $V_{\text{exact}}$ was the optimal value given by dynamic programming, and $V$ was the optimal value given by the particular algorithm. 

The first noteworthy result is the behavior of the exact $Z$ algorithm. Fig. \ref{fig:time_v1} and Fig. \ref{fig:time_v2} show that the exact $Z$ algorithm (yellow hexagon) is generally a fixed factor slower than the standard dynamic programming algorithm (blue square) for both types of instances, but has the same time scaling with $N$ and achieves similar accuracy (Fig. \ref{fig:v_n1} and Fig. \ref{fig:v_n2}). This reflects the fact that calculating the exact partition function for the KP is tantamount to implementing the standard dynamic programming solution to the KP, and thus both algorithms should proceed in $O(NW)$ time with more time required for the exact $Z$ algorithm in order to precisely compute its exponential factor.

The annealing algorithm (orange triangle) performs faster than the dynamic programming and exact $Z$ algorithms but more slowly than both large $W$ algorithms as $N$ increases. This time scaling reflects the fact that the state space annealing must explore grows as $2^N$ for increasing $N$ and thus takes longer to search for an optimal solution. The annealing algorithm is also less accurate than the algorithms for the strongly correlated (but not degenerate) circle instance, but achieves much better accuracy for the ratio-degenerate spanner instance. This likely reflects the fact that degenerate instances often allow for multiple solutions that an annealing algorithm can converge to, but less degenerate instances require a more extensive (and thus more error prone) search of the state space. 

For the large $W$ algorithms, we see that the nonzero-temperature algorithm (grey pentagon) is faster than the dynamic programming solution for values of $N \geq 10^2$. True to the logic of Fig. \ref{fig:accuracy_complexity}, the faster speed of the nonzero-temperature algorithm stems from the its  basis in solving an equation, a process which generally has a lower order scaling with $N$ than exact algorithmic solutions for the KP. The zero-temperature algorithm (red-plus) is even faster than the nonzero-temperature version because while the latter requires time consuming high precision solvers to compute exponential terms at low temperture, the former's constraint equation (\rfw{LWconstr}) has simple algebraic factors.

Considering the two large $W$ algorithm's relative accuracies, it is apparent that the nonzero-temperature algorithm is less accurate than the zero-temperature algorithm. We noted in Fig. \ref{fig:soln_temp_dependence} that the nonzero-temperature algorithm (Algorithm \ref{algo:tneq0}) becomes more accurate as $T$ is lowered, and thus nonzero values of $T$ will generally yield solutions that are not yet at their optimal possible values. Thus as the $T$ defining the nonzero-temperature algorithm is lowered, we expect the algorithm's predicted optimal value to converge to that of the zero-temperature algorithm (Algorithm \ref{algo:t=0}), provided there is a means to solve \rfw{GNz} for progressively lower $T$.

Although both large $W$ algorithms achieve high accuracies for the circle instance (which has strong correlations between values and weights) neither algorithm achieves high accuracies for the spanner instance (in which the value-to-weight ratio is highly degenerate across items). The reason for the poor performance stems from the properties of the spanner instance. For the chosen construction (with $\nu=2$ and $m=10$), there are only two possible values of the ratio $v_j/w_j$. The zero-temperature algorithm (Algorithm \ref{algo:t=0}) is blind to differences in values and weights where the ratio is the same, and accepts (or rejects) all items that fall above (or below) a given ratio. When there are two distinct ratios, the algorithm only has three choices of how to build up a solution: Accept all items with both ratios, accept all items of only one ratio and reject all items of the other, or reject all items with both ratios. Therefore, the algorithm does not exhibit the selectivity required to build up an optimal solution. This constrained set of choices leads to a poor accuracy for the zero-temperature algorithm. Moreover, for the given spanner instance, there are only $20$ possible unique choices for $(v_j, w_j)$ meaning that a system with $N\sim 1024$ items is highly-degenerate with on average 50 copies of each type of item. We recall from \rfw{degensoln}, that the zero-temperature algorithm yields ambiguous predictions for highly-degenerate instances, and so we would not expect it to perform well for the spanner instance. The nonzero-temperature version of the algorithm represents a softer (i.e., continuous) version of the acceptance criteria that governs the zero-temperature algorithm and thus shares zero-temperature algorithm's limitations.

The greedy algorithm is also dependent on the ratio $v_j/w_j$, but it is able to avoid the degeneracy problems of the spanner instance and achieve consistently accurate results because the algorithm adds items more selectively. In particular, the algorithm can continue to add items to the solution as long as doing so does not violate the weight constraint.
More generally, for both of the instance types, the greedy algorithm performs well for large values of $N$, eventually (for the largest $N$) predicting values close to the optimal total value in much less runtime. This might appear strange to those familiar with how greedy algorithms are discussed in KPs. The greedy algorithm is typically touted as a fast but inaccurate way to find an optimal collection of objects because it myopically looks at the next best choice rather than considering more holistic object combinations. However, given the large $W$ limit, the KP instances satisfy $W \gg w_{j}$ for all $j$, and thus including sub-optimal items generally does not preclude the inclusion of other items as needed. In other words, in the limit of $W\gg w_j$, the instances act as "semi-continuous KP" for which we expect the greedy algorithm to perfom well \cite{kellerer2004greedy}.

Comparing the performances of the three introduced algorithms to those of the standard algorithms for the KP, we see that the large $W$ algorithms generally do not perform better along the given metrics than existing algorithms. Such a result might strike one as indication of wasted effort. The statistical physics formalism has resulted in algorithms that are more complicated but yield no better performance than pedestrian KP algorithms. However, in Section \ref{sec:part_func}, we noted that the exact $Z$ algorithm was related to the dynamic programming solution to the KP. In the current section, given the definition of the standard KP greedy algorithm, we noted that it made use of value-to-weight ratios in a way similar to these ratios use in the zero-temperature algorithm. This relation in turn suggests that the zero-temperature algorithm can be related to a greedy algorithm. The fact that both the exact $Z$ and zero-temperature algorithms stem from a common statistical-physics starting point suggests that the established and conjectured counterparts of these algorithms (i.e., dynamic programming and the greedy algorithm, respectively) could be similarly related. Making this relationship concrete could in turn allow us to use statistical physics to connect distinct algorithmic spaces of combinatorial optimization problems. It is this connection, rather than the performances of the introduced algorithms, that gives the statistical physics perspective its value. We outline this connection more explicitly in the next section.

\section{Greedy Algorithm as Large $W$ Limit of Dynamic Programming \label{sec:greedy_dynamic}}


In this section, we show how the statistical physics formalism allows us to relate two approaches to solving the KP, ultimately revealing that the recursive DP solution leads to a myopic greedy solution when the problem is taken to the $W\gg w_j$ limit. 

First, we review the fundamental question asked by the zero-temperature and standard greedy algorithms.
The large $W$ zero-temperature algorithm (Algorithm \ref{algo:t=0}) asks "What is the minimum value-to-weight ratio (i.e., $\gamma_0$ defined in \rfw{gamma_lim}) above which one can accept all the satisfying items and obtain a total weight that satisfies the weight limit?" The standard KP greedy algorithm asks "Which items are included in the knapsack if one arranges all items in decreasing order of their value-to-weight ratios and accepts items in sequence until the knapsack has a total weight that satisfies the weight limit?"


In the greedy algorithm, the ordering and selective acceptance of items yields a minimum ratio for $v_{\ell}/w_{\ell}$ above which all objects are included knapsack.  Conversely, the zero-temperature algorithm seeks to compute a minimum acceptance-ratio as a first step and then to use this ratio as a rule for including items. In essence, the greedy algorithm follows an "ordering and fill-up procedure" while the the zero-temperature algorithm follows a "criteria procedure" for deciding which items are included in the knapsack, but both algorithms make use of a minimum value-to-weight ratio.

From here, we can ask whether the minimum-ratio calculated through the greedy algorithm is the same as the minimum ratio of the zero-temperature algorithm. To answer this question, we first define the minimum-ratio for the greedy algorithm. Following the standard greedy algorithm for the KP \cite{kellerer2004greedy}, say that the items in the knapsack are sorted so that their value-to-weight ratios are in non-increasing order: 
\begin{equation}
    \frac{v_1}{w_1} \geq \frac{v_2}{w_2} \geq \cdots \frac{v_N}{w_N}.
    \label{eq:greedy_ordering}
\end{equation}
Items are then added to the knapsack until the weight-limit is violated. We define $\alpha$ as the index of the item where this violation first occurs:
\begin{equation}
\alpha = \min\Big\{k : \sum_{j=1}^{k}w_j > W \Big\}.
\label{eq:alpha_def}
\end{equation}
By \rfw{greedy_ordering}, all items $j$ satisfying $v_j/w_j \geq v_{\alpha}/w_{\alpha}$ are inclueded in the greedy KP solution\footnote{Note that the converse is not true, namely not all items in the greedy solution must have value-to-weight ratios greater than $v_{\alpha}/w_{\alpha}$. Specifically, just because adding the $\alpha$th item violates the weight constraint does not mean adding any subsequent item will also violate the constraint.}, and thus $v_{\alpha}/w_{\alpha}$ is the desired minimum ratio. For the zero-temperature algorithm, the solution \rfw{Xalgsoln} indicates that $\gamma_0$ is the minimum value-to-weight ratio that determines item inclusion for the algorithm. This $\gamma_0$ value is found by selecting the $\gamma$ at which the equation
\begin{equation}
    L(\gamma; \tbf{v}, \tbf{w}, W) = W - \sum_{j=1}^{N} w_j H(v_j/w_j -\gamma_0)
\end{equation}
is zero or minimized for $L\geq 0$. From the minimum ratios of the greedy algorithm and the zero-temperature algorithm, we can define the "normalized ratio difference"  as 
\begin{equation}
    \text{norm. ratio diff.} \equiv \frac{|\gamma_0 - v_\alpha/w_\alpha|}{\gamma_0},
    \label{eq:ratio_diff}
\end{equation}
which defines the percent difference by which the greedy minimum-ratio differs from the large $W$ zero-temperature minimum-ratio. 
Computing the values of this difference for the circle and spanner instance, we find (Code for calculation is linked in Section \ref{sec:code}, \textit{Supplementary Code})
\begin{equation}
    \text{norm. ratio diff.} < 10^{-15},
\end{equation}
suggesting that the two minimum-ratio values are numerically identical. 

Thus, it appears that the standard greedy algorithm and the zero-temperature algorithm are implementing the same basic procedures under different framings. Still, whether that framing concerns ordering items and filling up the knapsack or is one based on a criteria, both the standard greedy algorithm and the zero-temperature algorithms reflect the type of local decision making that typifies a greedy algorithm. Both algorithms are seeking to optimize a variable (i.e., total value) by making choices that do not involve a full exploration of the state-space. Thus we can interpret the zero-temperature algorithm as a greedy algorithm.
\begin{equation}
    \begin{array}{ccc}
    \text{Greedy Algorithm} &  & \text{Large $W$ Zero-Temperature Algorithm} \\
     &\qquad \cong \quad &  \\
    (\textit{Fill knapsack up to min $v_{\ell}/w_{\ell}$ limit}) &  & (\textit{Compute min  $v_{\ell}/w_{\ell}$ limit, then fill knapsack}) \\
    \end{array}
\end{equation}

In so far as greedy algorithms go, the zero-temperature algorithm is a fairly unsophisticated one, sharing the basic characteristics of the standard greedy solution (namely a value-to-weight ratio and the fill to the weight limit procedure) without any of that algorithm's discernment. This difference is reflected in the two algorithm's relative performances in Fig. \ref{fig:spd_accy} and is explained by the fact that while the zero-temperature solution only includes items that are above a certain value-to-weight ratio, the standard greedy solution can accept items with ratios below this special ratio if doing so does not violate the weight limit.

Having concluded that the zero-temperature algorithm (Algorithm \ref{algo:t=0}) is a greedy algorithm, it appears that we have done a lot of work just to end up in essentially the same (and arguably a worse) place. If the large $W$ limit of the partition function for the KP yields a greedy algorithm, and this algorithm is less consistently accurate than the standard greedy algorithm, then what value does the new algorithm provide to computer scientists who are interested in finding better ways to solve real problems? Moreover, what value does this formalism provide to physicists who have already studied the statistical mechanics of this system as a representation of a disordered system \cite{korutcheva1994statistical, inoue1997statistical}? 
Here we argue that the value in both directions exists in the path we have taken to seemingly end up in the same place we began\cite{eliot1943little}. 

\begin{figure}[t]
    \centering
    \includegraphics[width=.7\linewidth]{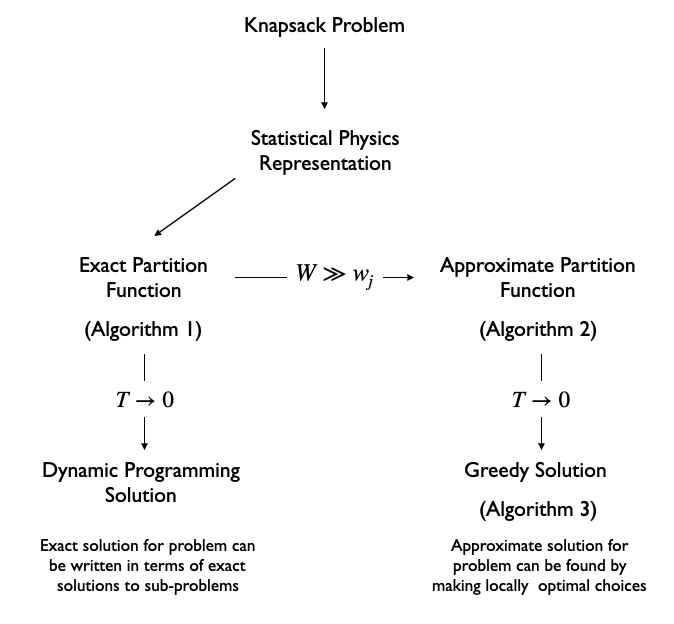}
        \caption{Relationship between Dynamic Programming and Greedy Solutions to the KP: For the statistical physics representation of the knapsack problem, taking the $T\to 0$ limit of the exact partition function yields the standard dynamic programming solution, and taking the $T\to0$ limit of the large-$W$ limit
        of the partition function yields a greedy solution (Algorithm \ref{algo:t=0}). The parameter $T$ modulates how far away we are from the optimal solution (i.e., lower $T$ corresponds to more optimal solutions). Thus both Algorithm \ref{algo:exactz} and Algorithm \ref{algo:tneq0} represent, respectively, dynamic and greedy solutions to the KP.  In this way, the large $W$ limit of the dynamic programming solution is a greedy solution.}
    \label{fig:algo_relat}
    \end{figure}

Given the partition function identity in \rfw{dynamic_prog} and the prior discussion on how the zero temperature algorithm is a greedy algorithm, the statistical physics formalism suggests that there is a relationship between dynamic programming and greedy algorithms, at least for the KP. In combinatorial optimization, dynamic programming and greedy approaches to solving problems are by definition very different. In dynamic programming, a problem is solved exactly by recursively referring to stored solutions of sub-problems.  In greedy algorithms, a (typically approximate) solution is built by selecting whatever choice is best given an initial state. After deriving the identity \rfw{dynamic_prog}, we showed that the exact partition function \rfw{knapsack_prelim} yielded the dynamic programming solution to the KP, and above, we explained how the zero-temperature large $W$ algorithm was a greedy algorithm. From the fact that we can move from the exact partition function to the approximate partition function by taking the large $W$ limit, in a sense, applying a large $W$ approximation to the dynamic programming approach to the KP yields a greedy algorithm. We represent the relationships between these algorithms and limits in Fig. \ref{fig:algo_relat}. 


The merit in establishing this relationship can be understood through an analogy. One could imagine knowing that the quantity $I_N = \ln(N!)$ can be calculated recursively as $I_N = \ln N +I_{N-1}$ and also knowing that $I_N$ can be approximated for large $N$ as $I_N \simeq N \ln N -N$, but \textit{not} knowing how the approximation relates to the exact result or why the approximation is a good one. To make clear the relationship between the exact and approximate forms of $I_N$, we could use the formalism of the Gamma function plus Laplace's method (or alternatively just the Euler-Maclaurin formula) to derive the latter from the former and in turn obtain a systematic way to obtain higher-order corrections to the approximation. As an added benefit, the derivation would serve as an example for computations of similar combinatorial factors, thus providing many other connections between exact and approximate combinatorial expressions. 

The statistical physics model for the KP achieves something similar for the dynamic programming and greedy solutions to the KP. It shows how the two algorithms are related and serves as an example for how similar ideas could be applied to find relationships between other algorithms. This mapping only arises from the statistical physics framing of the system, and the representation of the algorithms themselves is specific to the KP, but its existence gives us a more comprehensive understanding of how algorithms for the KP relate to one another and motivates further exploration into whether this relationship generalizes for other similarly formulated combinatorial optimization problems.

\section{Discussion \label{sec:discuss}}

Starting from \reffig{accuracy_complexity}, we began this work by suggesting that the properties of analytical statistical physics made the formalism amenable to solving large $N$ combinatorial optimization problems with high accuracy and in less time than exact solutions. From \reffig{spd_accy}, this suggestion appears to have supporting evidence in some cases, but it is also clear that already well-known algorithms outperform the ones introduced in this work. 

We explored three related algorithms originating from the statistical physics formulation of the problem. The first algorithm (Algorithm \ref{algo:exactz}) followed from a recursive identity for the exact partition function of the system. This algorithm was found to reproduce the standard dynamic programming solution for the KP. Next, by approximating the exact partition function in the large $W$ limit, we obtained an approximate solution to the KP which resulted in another algorithm (Algorithm \ref{algo:tneq0}) that was termed the "nonzero-temperature algorithm." Finally, by taking this latter solution to the $T\to0$ limit, we obtained the "zero-temperature algorithm" (Algorithm \ref{algo:t=0}). 

Comparing the accuracies and runtimes of these algorithms to the standard dynamic programming algorithm, the standard greedy algorithm, and simulated annealing, we found that the introduced algorithms did not consistently outperform existing ones for "difficult instances" (\reffig{spd_accy}). Algorithm \ref{algo:exactz} had accuracy results similar to those of the standard DP algorithm, but was a fixed factor slower. Algorithm \ref{algo:tneq0} generally performed worse than Algorithm \ref{algo:t=0}, and while Algorithm \ref{algo:t=0} did perform better as $N$ increased (with not as much of a runtime increase as that for the exact algorithm), it did not perform well for degenerate KP instances, and the standard greedy algorithm always had better results. 

Since well-known algorithms do better in runtime and accuracy than the introduced algorithms, it may seem that the new algorithms have little value. However, the path leading to the approximate algorithms suggests that the value of the statistical physics approach exists outside of what initially motivated it. In particular, the correspondence between the statistical physics algorithms and the more well-known algorithms tells us something about the relationships between the algorithms themselves.

Dynamic programming and greedy solutions to a combinatorial optimization problem are typically seen as distinct and separate ways to solve a common problem. Upon inspecting the two algorithms, it does not appear that they are at all related other than the fact that they use the same parameters. However the organization of Fig. \ref{fig:algo_relat} reveals that when the KP is formulated as a statistical physics system, the exact partition function yields the standard dynamic programming solution, and the large $W$ limit of that partition function yields a greedy solution to the KP. Thus, through the statistical physics representation of the KP, we see that taking the large $W$ limit of a dynamic programming solution to the KP yields a greedy solution (Algorithm \ref{algo:t=0}). 

In a way, it makes sense that the large $W$ approximation to the KP partition function would yield a greedy algorithm. The large $W$ approximation is based on the method of steepest descent which approximates the partition function by the local minimum of a potential function. The search for this local minimum can be represented as a directed step-by-step movement towards the minimum of the potential, essentially as a greedy search. So the large $W$ approximation already contains a greedy solution within its structure. What is interesting here is that the greedy approach to maximizing the continuous potential is paralleled by a greedy approach for the final discrete KP solution.


There are a few investigations that could naturally follow this work: Understanding error estimates of the solution; finding higher-order corrections to the approximate solutions; applying the formalism to other optimization problems; 

The algorithms give us solutions for the KP in the form of the vector $\tbf{X}$, but they do not give us a sense for how the approximated maximum $\tbf{v}\cdot \tbf{X}$ differs from the true maximum. For the exact $Z$ algorithm, such an error estimate could follow from a $T\ll 1$ expansion of the partition function and the argument of the limit in \rfw{XsolnSimp}. For the large $W$ algorithms, this error estimate could likely be found by a more careful accounting of the higher order terms in the steepest descent approximation. 

These higher-order terms could also serve as the foundation for improved versions of Algorithm \ref{algo:tneq0} and \ref{algo:t=0}. In the standard steepest descent approximation, the potential of the exponential integrand is approximated by a quadratic function. It is this function which yielded the approximate solution \rfw{kn_soln} from which the final solutions \rfw{xellsoln} and \rfw{Xalgsoln} were derived. By retaining higher-order terms in the integrand expansion, we could obtain more accurate partition functions and in turn more accurate approximate solutions, possibly ones that do not have the all-or-nothing deficiencies of the existing large $W$ algorithms. 

Following the formalism that led to Algorithms \ref{algo:exactz}, \ref{algo:tneq0}, and \ref{algo:t=0}, it is clear that the ability to relate various limiting forms of the algorithms to each other follows from the representation of the KP partition function as the integral \rfw{knapsack_prelim}. The integral representation of the KP allowed us to apply the steepest descent approximation in the large $W$ limit, which ultimately connected the dynamic programing representation and the greedy representation of the problem. This integral representation in turn followed from representing the KP weight-constraint as a contour integral. Thus, it would be straightforward to extend this approach to other combinatorial optimization problems with similar types of constraints. In particular, if a problem is restricted by an inequality constraint (leading to a Heaviside function) or an equality constraint (leading to a Kronecker delta function), then much of the prior formalism could be transferred wholesale, and thus repeated application of the formalism to different optimization problems could lead to many correspondence pairs between various dynamic programming and greedy solutions to problems. Likely a general theorem could be established for this class of problems, that states that the dynamic programming solutions to all combinatorial optimization problems of a certain type yield a greedy solution in a certain limit. One necessary limitation to such extensions is that in order for an integral approximation to be possible, the number of constraints cannot scale linearly with $N$ \cite{shun1995laplace}, and thus problems like the Traveling Salesperson Problem would be resistant to this approach. 

In all, this work represents a new way to frame the value of statistical physics to combinatorial optimization. For many decades, computational statistical physics algorithms have been used to stochastically search for the solutions to combinatorial optimization problems. And combinatorial optimization problems themselves have been used to study the properties of rugged landscapes in statistical physics systems. In contrast to these previous approaches, the main value of the current work is not a new highly performant algorithm, or a new context in which to apply the replica-method, but a potentially new way to understand the relationships between solutions to combinatorial optimization problems with constraints. Greedy algorithms are known ways to approximate the solutions to combinatorial optimization problems. These approaches are understood as distinct from dynamic programming which are grounded in recursive solutions to a problem. Here we have shown how each can be represented as a branch from the common starting point of the statistical physics formulation of the KP.

\section{Conclusions}
We represent the Knapsack Problem (KP) as a statistical physics system and use the representation to obtain three algorithms for the KP. The relationships between the algorithms reveal relationships between well-known algorithms for the KP. In particular, taking a large $W$ limit of a dynamic programming solution to the KP yields a greedy solution to the KP. 

\section{Acknowledgements}
This work was completed in Jellyfish. The author thanks Nicholas Arcolano, Arlo Clarke, Amir Bitran, Rostam Razban, Michael Brenner, and Pankaj Mehta for useful comments.

\section{Funding}
This research received no specific grant from any funding agency, commercial, or nonprofit organization.

\section{Conflicts of Interest}
The author has no conflicts of interest to disclose.

\section{Data Availability Statement \label{sec:code}}
The numerical work was done with a 2.2 GHz 6-Core Intel Core i7 processor using Python 3.9 to computationally implement all the algorithms. Code used to generate all figures is found in the repository \href{https://github.com/mowillia/LargeWKP}{https://github.com/mowillia/LargeWKP}.

\appendix

\section{Dynamic Programming KP from Partition Function \label{app:dynamic}}

In this section, we show that the standard dynamic programming solution to the KP is contained within the exact KP partition function. We recall that the KP partition function 
\begin{equation}
Z_{N}(\beta \tbf{v}, \tbf{w}, W) = \sum_{\tbf{x}} \Theta\Big( W- \tbf{w}\cdot \tbf{x} \Big) \exp\Big(\beta \tbf{v}\cdot \tbf{x} \Big),
\label{eq:knapsack_pf0}
\end{equation}
has the associated identity
\begin{equation}
Z_N(\beta \tbf{v}, \tbf{w}, W)  = Z_{N-1}^{(N)}(\beta \tbf{v}, \tbf{w}, W) + e^{\beta v_N}Z^{(N)}_{N-1}(\beta \tbf{v}, \tbf{w}, W-w_N),
\label{eq:dynamic_prog0}
\end{equation}
where $Z^{(N)}_{N-1}(\beta \tbf{v}, \tbf{w}, W)$ is the partition function in which the $N$th component is eliminated from both $\tbf{v}$ and $\tbf{w}$, and thus only $N-1$ items are under consideration. 

Using \rfw{knapsack_pf0}, the average value $\langle V \rangle_{N, W} \equiv \langle \tbf{v} \cdot \tbf{x}\rangle_{N,W}$ can be written in terms of the $\beta$ derivative of the partition function:
\begin{equation}
\langle V \rangle_{N, W} = \frac{1}{Z_N(\beta \tbf{v}, \tbf{w}, W)}\frac{\partial}{\partial \beta} Z_N(\beta \tbf{v}, \tbf{w}, W)
\label{eq:VNW}
\end{equation}
where subscripts $N, W$ signify that the average is defined for a weight limit $W$ with $N$ items under consideration. Differentiating \rfw{dynamic_prog0} with respect to $\beta$, dividing the result by $Z_N(\beta \tbf{v}, \tbf{w}, W)$, and using \rfw{VNW} then yields
\begin{equation}
\langle V \rangle_{N, W} =\frac{ \partial_{\beta}Z_{N-1}^{(N)}(\beta \tbf{v}, \tbf{w}, W) + e^{\beta v_N}(v_{N} + \partial_{\beta})Z^{(N)}_{N-1}(\beta \tbf{v}, \tbf{w}, W-w_N)}{Z_{N-1}^{(N)}(\beta \tbf{v}, \tbf{w}, W) + e^{\beta v_N}Z^{(N)}_{N-1}(\beta \tbf{v}, \tbf{w}, W-w_N)}
\label{eq:VNWred}
\end{equation}

We know that given the definition of $\langle V \rangle_{N, W}$ as a statistical physics average and $V_{N}(W)$ as the maximum total-value of the corresponding optimization problem, we have the equality
\begin{equation}
V_{N}(W) = \lim_{\beta\to\infty} \langle V \rangle_{N, W}.
\label{eq:VNW_lim}
\end{equation}
And so, taking the limit of \rfw{VNWred} as $\beta \to \infty$, we obtain
\begin{align}
V_{N}(W) 
& = \lim_{\beta\to \infty} \frac{\langle V\rangle_{N-1, W} + \lambda_{N, W}(v_{N} + \langle V\rangle_{N-1, W-w_{N}} )}{1 + \lambda_{N, W}}
\label{eq:VNW2}
\end{align}
where we used \rfw{VNW} in the final equality, and we defined 
\begin{equation}
\lambda_{N, W} \equiv \frac{e^{\beta v_N}Z^{(N)}_{N-1}(\beta \tbf{v}, \tbf{w}, W-w_N)}{Z_{N-1}^{(N)}(\beta \tbf{v}, \tbf{w}, W) }.
\label{eq:lambda_def}
\end{equation}

In order to compute the right hand side of \rfw{VNW2}, we first note that there are three possible cases in which the limit can be evaluated: The case where $W< w_N$; the case where $W\geq w_N$ with the numerator of \rfw{lambda_def} dominating the denominator for $\beta \to \infty$; the case where $W\geq w_N$ with the denominator of \rfw{lambda_def} dominating the numerator for $\beta \to \infty$. 

For the case where $W< w_N$, the second term in \rfw{dynamic_prog0} (or, equivalently, the second term in the numerator or denominator of \rfw{VNW2}) vanishes because the partition function $Z_N$ cannot be defined for negative weight. Thus, we find $\lambda_{N, W} =0$ for $W< w_N$, and applying the definition \rfw{VNW_lim} to the right hand side of \rfw{VNW2}, we find
\begin{equation}
V_{N}(W)  = V_{N-1}(W) \qquad \text{[for $W< w_N$]}.
\label{eq:case1}
\end{equation}

For the next two cases defined by $W\geq w$, we recall that the partition function is a Boltzmann weighted sum over states where the argument of the Boltzmann weight is proportional to the total value of the collection for the corresponding state. When we take $\beta\to\infty$, the differences in the total values between states become more pronounced so that the partition function effectively becomes defined by its maximum total-value state. Specifically, in this limit, a general partition function $Z$ becomes $Z = e^{\beta {\cal O}_{\text{max}}} + \cdots$ where ${\cal O}_{\text{max}}$ is the maximum total-value and "$\cdots$" stands for sub-leading terms in this limit.

Thus whether the  numerator of \rfw{lambda_def} dominates the denominator in the $\beta \to \infty$ limit is entirely dependent on the relative values of the maximum total-values associated with each partition function: If the maximum total-value for $Z_{N-1}^{(N)}(\beta \tbf{v}, \tbf{w}, W) $ is greater than that for $e^{\beta v_N}Z^{(N)}_{N-1}(\beta \tbf{v}, \tbf{w}, W-w_N)$ then $\lim_{\beta \to \infty} \lambda_{N, W}  = 0$. Alternatively, if the maximum total-value for $Z_{N-1}^{(N)}(\beta \tbf{v}, \tbf{w}, W) $ is less than that for $e^{\beta v_N}Z^{(N)}_{N-1}(\beta \tbf{v}, \tbf{w}, W-w_N)$ then $\lim_{\beta \to \infty} \lambda_{N, W}  = \infty$. For the partition function $Z_{N-1}^{(N)}(\beta \tbf{v}, \tbf{w}, W)$, the maximum total-value is by definition $V_{N-1}(W)$, and for the partition function $e^{\beta v_N}Z^{(N)}_{N-1}(\beta \tbf{v}, \tbf{w}, W-w_N)$, the maximum total-value is $v_N + V_{N-1}(W-w_N)$ where we have $v_N$ as a first term since the factor $e^{\beta v_N}$ adds a value-element of $v_N$ to each state. 

Having determined the maximum total-values associated with the numerator and denominator of \rfw{lambda_def}, and given that the relative values of these maxima determine whether $\lambda_{N, W}\to 0$ or $\to \infty$ in the $\beta\to \infty$ limit, we find that \rfw{VNW2} (for $W\geq w_N$) becomes
\begin{equation}
V_{N}(W)  = V_{N-1}(W) \qquad \text{[for $ V_{N-1}(W) > v_N + V_{N-1}(W-w_N) $]},
\label{eq:case2}
\end{equation}
and the opposite case (still with $W\geq w_N$) yields
\begin{equation}
V_{N}(W)  = v_N + V_{N-1}(W-w_N) \qquad \text{[for $ V_{N-1}(W) < v_N + V_{N-1}(W-w_N) $]}.
\label{eq:case3}
\end{equation}
Reviewing the cases \rfw{case1}, \rfw{case2}, and \rfw{case3}, we see that together they produce the solution 
\begin{equation}
V_{N}(W) = \begin{dcases} V_{N-1}(W) & \text{for $W< w_N$} \\ \max\{V_{N-1}(W),\, v_{N} + V_{N-1}(W-w_N)\} & \text{for $W\geq  w_N$},\end{dcases}
\label{eq:bellman0}
\end{equation}
where $V_{j}(Q)$ is the maximum total-value of the knapsack with the first $j$ items included and a weight-limit $Q$. \rfw{bellman0} is the standard dynamic programming solution to the KP and we can thus conclude that the KP partition function reproduces the standard dynamic programming solution to the KP.

\section{Variations \label{app:var}}
In this appendix, we discuss the various generalizations of the 0-1 KP: The bounded, unbounded, and continuous \cite{kellerer2004general} KPs. All generalizations keep the essential problem format of \rfw{knap_constr} except each changes the state space available to each object and consequently also changes the associated sum over states. This change in summation alters the solubility of the final partition function and our ability to approximate it. We find that although we can develop algorithms analogous to those found in the main text for the bounded and unbounded KPs, we can only write down the partition function for the continuous KP. 

\subsection{Bounded Knapsack Problem \label{app:bound}}

Here we sketch the large $W$ algorithm for the bounded version of the KP. We focus on writing the results since the relevant derivations are largely identical to those for the 0-1 problem.

In the bounded KP, we allow each object $i$ to appear a maximum of $C_i$ times in the final collection. We collectively represent these maximum constraints through the vector $\tbf{C} = (C_1, C_2, \ldots, C_N)$. For the statistical-physics representation of the problem, this change amounts to replacing the summation \rfw{01summ} with 
\begin{equation}
\sum_{\tbf{x}} \equiv \prod_{j=1}^N \sum_{x_{j} =0}^{C_j} \qquad \text{[Summation for "multiple copies" problem]},
\label{eq:Msumm}
\end{equation}
where $x_{i}$ denotes the number of times object $i$ is included in the final collection. Following a derivation similar to that in Sec. \ref{sec:part_func}, we find that the partition function for this system is 
\begin{equation}
Z_{N}\left(\beta \tbf{v}, \tbf{w}, \tbf{C}, W\right) = \frac{1}{2\pi i} \oint_{\Gamma} \frac{dz}{z^{W+1}} \frac{1}{1-z} \prod_{k=1}^N \frac{1-z^{(C_k+1)w_k}e^{(C_k+1)\beta v_k}}{1-z^{w_k}e^{\beta v_k}}.
\label{eq:ZNC}
\end{equation}
For this partition function, we can derive a recursive relation analogous to \rfw{dynamic_prog} and ultimately connect \rfw{ZNC} to the dynamic programming solution to the bounded KP. Noting that 
\begin{equation}
 \frac{1-z^{(C_N+1)w_N}e^{(C_N+1)\beta v_N}}{1-z^{w_N}e^{\beta v_N}} = 1 + z^{w_N} e^{\beta v_N}  \frac{1-z^{C_Nw_W}e^{C_N\beta v_N}}{1-z^{w_N}e^{\beta v_N}}
\end{equation}
we find that \rfw{ZNC} implies
\begin{equation}
Z_{N}\left(\beta \tbf{v}, \tbf{w}, \tbf{C}, W\right) = Z^{(N)}_{N-1}\left(\beta \tbf{v}, \tbf{w}, \tbf{C}, W\right) + e^{\beta v_N} Z_{N}\big(\beta \tbf{v}, \tbf{w}, \tbf{C}^{(N)}, W-w_N\big)
\label{eq:dynamic_prog_bound}
\end{equation}
where $Z_{N-1}^{(k)}\left(\beta \tbf{v}, \tbf{w}, \tbf{C}, W\right)$ is the partition function \rfw{ZNC} where the $k$th component is eliminated from all the vectors and thus only $N-1$ components are under consideration, and we defined $\tbf{C}^{(k)} \equiv (C_1, \ldots, C_k-1, \ldots, C_N)$. Letting $V_{k}(W, \tbf{C})$ be the optimal value for the bounded KP instance where only the first $k\leq N$ items are used, we can show (using arguments similar to those Appendix \ref{app:dynamic}) that \rfw{dynamic_prog_bound} implies
\begin{equation}
V_{N}(W, \tbf{C}) = \begin{dcases} V_{N-1}(W, \tbf{C}) & \text{for $W < w_N$} \\  \max \{V_{N-1} (W, \tbf{C}),\, v_N + V_{N}(W-w_N, \tbf{C}^{(N)}) \} & \text{for $W  \geq w_N$} \end{dcases},
\end{equation}
which defines the dynamic programming algorithm for the bounded KP. 

With \rfw{ZNC}, we can also apply the method of steepest descent (as in Sec. \ref{sec:largeN}) to approximate this partition function and derive the expressions needed to formulate algorithms akin to those in Sec. \ref{sec:soln}.

For these new algorithms, we primarily need new expressions for \rfw{GNz},  \rfw{xellsoln}, and \rfw{Xalgsoln}. Deriving an expression analogous to \rfw{GNz} is straightforward; the calculation is identical to that for the 0-1 problem. We find 
\begin{equation}
G_{N}\left(z; \beta \tbf{v}, \tbf{w}, \tbf{C}, W\right) = -W + \frac{z_0}{1-z_0}  -\sum_{i=1}^{N}\frac{w_{i}}{1-e^{-v_i/T} z_0^{-w_i} }+\sum_{i=1}^{N}\frac{(C_i+1)w_{i}}{1-e^{-(C_i+1)v_i/T} z_0^{-(C_i+1)w_i}}.
\label{eq:GNzC}
\end{equation}
Similarly, the expression for the analog of \rfw{xellsoln} can be found by differentiating the potential term inferred from \rfw{ZNC}. From this differentiation we find
\begin{equation}
\langle x_{\ell} \rangle  = \frac{C_\ell+1 }{1-e^{-(C_\ell+1)v_\ell/T}z_0^{-(C_\ell+1)w_\ell} }-\frac{ 1}{1-e^{-v_\ell/T}z_0^{-w_\ell}} + {\cal O}(w_j/W).
\label{eq:xellsolnC}
\end{equation}
From here, our formulation of the algorithm can proceed in one of two directions to obtain $X_{\ell}$: We can take the $T\to0$ limit of \rfw{xellsolnC}; or we can consider \rfw{xellsolnC} with a finite threshold that allows us to determine the assignment of objects. 

Following the first path, we consider the function
\begin{equation}
f_{B}(x; \alpha) = \frac{B+1}{1- e^{-(B+1)x/\alpha}} - \frac{1}{1-e^{-x/\alpha}}.
\end{equation}
Combining each term under a common denominator and applying L'Hopital's rule twice yields $f_{B}(0; \alpha) = B/2$. Taking $\alpha \to 0$, we then find $\lim_{\alpha\to 0} f_{B}(x; \alpha) = B H(x)$ where $H(x)$ is the Heaviside step function that is defined at zero by $H(0) = 1/2$. Therefore, we see that the $T\to 0$ limit of \rfw{xellsolnC} is \vspace{.25cm}
\begin{equation}
X_{\ell} = \lim_{T\to 0} \langle x_{\ell} \rangle = C_{\ell} H(v_{\ell}-\gamma_0 w_{\ell})  + {\cal O}(w_j/W)., \vspace{.35cm}
\label{eq:XsolnC}
\end{equation}
where $\gamma_0 = \lim_{T\to 0}\gamma(T)$, $\gamma(T) \equiv T \ln z_0(T)$, and $z_0(T)$ is the solution obtained from the constraint condition \rfw{GNzC} being set to zero. \rfw{XsolnC} allows us to formulate the $T=0$ algorithm for the bounded KP. 

For the second path, we have to use a threshold that can convert $\langle x_\ell\rangle$ to an integer. We introduce this threshold so that we can round $\langle x_{\ell}\rangle$ to the nearest integer based on how stringent we want to be about the rounding. To this end, we will define a $p_{\text{thresh}}\in (0, 1)$ and define the solution $X_{\ell}$ as 
\begin{equation}
X_{\ell} = \begin{dcases} C_{\ell} & \text{if $\langle x_{\ell} \rangle> C_{\ell} \,p_{\text{thresh}}$} \\ 0 & \text{otherwise.}\end{dcases}
\label{eq:XCdef}
\end{equation}
In \rfw{XCdef}, we have forced the occupancy for object $\ell$ to either be zero or its maximum number of instances in the collection. We chose this "all or nothing" decision criterion in order to have a solution similar to that for \rfw{Xelldef}. Such a choice also well corresponds with the $T=0$ solution \rfw{XsolnC}.


\subsection{Unbounded Knapsack Problem}

Here we sketch the large $W$ algorithm for the unbounded version of the KP. We again focus on writing the results since the relevant derivations are largely identical to those for the 0-1 problem.

In the unbounded KP, we allow each object $i$ to appear a countably infinite number of times in the final collection. For our statistical-physics representation of the problem this change amounts to replacing the summation \rfw{01summ} with 
\begin{equation}
\sum_{\tbf{x}} \equiv \prod_{j=1}^N \sum_{x_{j} =0}^{\infty} \qquad \text{[Summation for "infinite" problem]},
\label{eq:Infinitesumm}
\end{equation}
where $x_{i}$ denotes the number of times object $i$ is included in the final collection. If we were to follow a derivation similar to that in Sec. \ref{sec:part_func}, we would reach the expression 
\begin{equation}
Z_{N, \infty}\left(\beta \tbf{v}, \tbf{w},  W\right) = \frac{1}{2\pi i} \oint_{\Gamma} \frac{dz}{z^{W+1}} \frac{1}{1-z} \prod_{k=1}^N\sum_{x_k = 0}^{\infty} \left(z^{w_k} e^{\beta v_k} \right)^{x_k}. 
\label{eq:ZNinfty0}
\end{equation}
This expression contains an infinite series, and for such series there are conditions on whether the final result is finite. In this case, in order for \rfw{ZNinfty0} to be finite we require
\begin{equation}
|z| < e^{- \beta v_k/w_k} \quad \text{for $k=1, 2, \ldots, N$}.
\label{eq:zcondition}
\end{equation}
We can ensure this condition by choosing a contour $\Gamma_R$ whose $z$ values satisfy $z \in R$ where $R$ is the circle in the complex plane with radius equal to $\min (e^{- \beta v_1/w_1},e^{- \beta v_2/w_2}, \ldots, e^{- \beta v_N/w_N})$. 

Constraining our contour in this way, we find that the partition function is 
\begin{equation}
Z_{N, \infty}\left(\beta \tbf{v}, \tbf{w},  W\right) = \frac{1}{2\pi i} \oint_{\Gamma_R} \frac{dz}{z} \exp F_{N, \infty}\left(z; \beta \tbf{v}, \tbf{w},  W\right),
\label{eq:ZNinfty}
\end{equation}
where we defined 
\begin{equation}
F_{N, \infty}\left(z;\beta \tbf{v}, \tbf{w},  W\right) \equiv  - W \ln z - \ln (1-z) - \sum_{k=1}^N\ln \left(1- z^{ w_k}e^{\beta v_k}\right)
\label{eq:FNinfty}
\end{equation}

With \rfw{ZNinfty}, we can apply the method of steepest descent as in Sec. \ref{sec:largeN} to approximate this partition function and derive the necessary expressions to formulate algorithms akin to those for the 0-1 problem in Sec. \ref{sec:soln}. Namely, we can derive expressions analogous to \rfw{GNz} and \rfw{xellsoln}, or \rfw{GNzC} and \rfw{xellsolnC}

However, the $z$ value at which $F_{N, \infty}$ is minimized is not always included within the region of contours for which the infinite series that takes us from \rfw{ZNinfty0} to \rfw{ZNinfty} is valid. In other words, the $z$ that minimizes \rfw{FNinfty} can violate \rfw{zcondition} and thus exist off the contour $\Gamma_R$. In such a case, we cannot sensibly evaluate the integrand of the partition function at this minimum potential value. Therefore, the method of steepest descent cannot always be applied to the partition function in the unbounded case.

We can rectify this by noting that no non-trivial KP is truly unbounded: If each item has a weight, there must be a maximum number of each item that can be included in the knapsack. Thus we can convert the unbounded KP into a bounded KP where the bound is defined by 
\begin{equation}
C_{\ell} \equiv \floor{W/w_{\ell}}.
\label{eq:Cdef}
\end{equation}
Thus, the large $W$ algorithms for the unbounded case are identical to the algorithms sketched in Appendix \ref{app:bound} except that the components of  $\tbf{C}$, rather than being a set of independently defined problem parameters, are computed from the weights and weight-limit through \rfw{Cdef}.

\subsection{Continuous Knapsack Problem}
Unlike the bounded and unbounded KPs, the continuous KP has a continuous space of states. Consequently, the previous discrete delta and Heaviside function methods do not apply, and we cannot reach an expression to which we can apply the method of steepest descent. Therefore, it seems that we cannot obtain a statistical-physics-based algorithm for this case. Still, it is possible to achieve the limited victory of writing a partition function for this problem and we do so here.

In the continuous KP, $x_i$ can take on any real number between $0$ and $1$ inclusive, and each $x_i$ represents the amount of object $i$ we include in the final collection. We take $\tbf{x} = (x_1, x_2, \ldots, x_N)$ to represent the amounts of each object in the collection, and the total weight and total value of the objects to be $\tbf{w} \cdot \tbf{x}$ and $\tbf{v}\cdot \tbf{x}$, respectively, where $\tbf{w} = (w_1, w_2, \ldots, w_N)$ and $\tbf{v} = (v_1, v_2, \ldots, v_N)$ are the corresponding weight and value vectors. 

Generalizing our previous discrete-space analysis to this continuous-space case amounts to a change from a discrete sum to a continuous integration. The partition function for this continuous case is
\begin{equation}
Z^{\text{cont}}_{N}(\beta\tbf{v}, \tbf{w},  W) = \int^{1}_{0} d^N\tbf{x}\, \overline{\Theta}\left( W- \tbf{w}\cdot \tbf{x} \left) \,\exp\right(\beta \tbf{v} \cdot \tbf{x} \right),
\label{eq:knapsack_cont}
\end{equation}
where
\begin{equation}
\int^{1}_{0} d^N\tbf{x} \equiv \int^{1}_{0} dx_1 \int^{1}_{0}dx_2 \ldots\int^{1}_{0} dx_N
\end{equation}
is the integration over the state space and where 
\begin{equation}
\overline{\Theta}(x) = \begin{dcases} 1 & \text{if $x \geq 0$} \\ 0 & \text{otherwise}, \end{dcases}
\end{equation}
is the Heaviside step function with a continuous argument. With such a continuous argument, the step function has the integral representation
 \begin{equation}
\overline{\Theta}(x) = \frac{1}{2\pi i} \lim_{\ve \to 0^+}\int^{\infty}_{-\infty} dk \, \frac{e^{i kx}}{k-i \ve},
\end{equation}
which can be affirmed by transforming the real-space integral in $k$ into a closed contour integral in the upper-half of the complex plane. Inserting this expression into \rfw{knapsack_cont}, we obtain 
\begin{align}
Z^{\text{cont}}_{N}(\beta\tbf{v}, \tbf{w},  W) & = \int^{1}_{0} d^N\tbf{x}\, \frac{1}{2\pi i} \lim_{\ve \to 0^+}\int^{\infty}_{-\infty} dk \, \frac{e^{i k(W- \tbf{w}\cdot \tbf{x})+ \beta \tbf{v}\cdot \tbf{x}}}{k-i \ve}\mm
 & = \frac{1}{2\pi i} \lim_{\ve \to 0^+}\int^{\infty}_{-\infty} dk \,\frac{e^{i kW}}{k-i \ve} \int^{1}_{0} d^N\tbf{x}\, e^{\tbf{x}\cdot (\beta \tbf{v} - i k\tbf{w})}\mm
 & =  \frac{1}{2\pi i} \lim_{\ve \to 0^+}\int^{\infty}_{-\infty} dk \,\frac{e^{i kW}}{k-i \ve} \prod_{j=1}^{N} \int^{1}_{0} dx_j \, e^{x_j(\beta v_j - i k w_{j})}, 
\end{align}
ultimately yielding the partition function 
\begin{equation}
Z^{\text{cont}}_{N}(\beta\tbf{v}, \tbf{w},  W)  = \frac{1}{2\pi i} \lim_{\ve \to 0^+}\int^{\infty}_{-\infty} dk \,\frac{e^{i kW}}{k-i \ve} \prod_{j=1}^{N} \frac{e^{(\beta v_j - i k w_{j})}-1}{\beta v_j - i k w_{j}}.
\label{eq:ZNcont}
\end{equation}
Since we are no longer integrating over a contour in the complex plane, applying a large $W$ approximation scheme to \rfw{ZNcont}  would require a slightly different formalism from that presented in this paper. Interestingly, fast approximation schemes are not needed for the continuous KP as it can be solved exactly via an $O(N\log N)$ greedy algorithm \cite{dantzig1957discrete} and thus what would typically be the fastest "approximation" method actually yields the exact solution.

\section{Additional Instances \label{app:additional_instances}}

We consider the runtime and accuracy results for the two other special instances introduced in \cite{pisinger2005hard}: The "profit-ceiling instance" and the "multiple strongly correlated Items" instance.

\begin{itemize}

    \item \tbf{Profit-Ceiling Instances:} Values are multiples of a given integer parameter $d$. The instance is generated by randomly (and uniformly) selecting weights $w$ in the interval $[1,R]$, for a free integer parameter $R$, and then setting the values to $p_j = d \lceil w_j/ d\rceil$. \cite{pisinger2005hard} noted that particularly difficult instances appeared by choosing $d=3$. The value of $R$ was set to  $100$. 
    
    \item \tbf{Multiple Strongly Correlated Items Instances:} Values are related to weights through $v = w + k_i$ where $k_i$ for $i=1, 2$ for the case of two strongly correlated instances. Specifically, the instances are generated as follows: The weights of $N$ items are randomly distributed in $[1, R]$. If the weight $w_j$ is divisible by a chosen $d$, we set the value $v_j = w_j + k_1$, otherwise we set it to $v_j = w_j + k_2$. 
    
   In accordance with \cite{pisinger2005hard}, we set $d=6$, $k_1 = 3R/10$, and $k_2 = 2R/10$. The value of $R$ was set to $100$. 

    \end{itemize}

    \begin{figure*}[t!]
        \centering
        \begin{subfigure}[t]{0.32\textwidth}
            \centering
            \includegraphics[width = \linewidth]{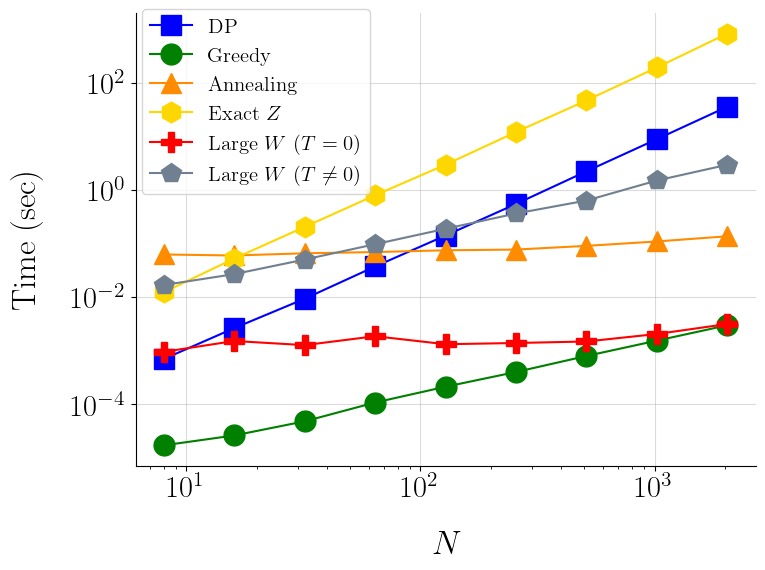}
            \caption{Profit-Ceiling: Time vs. $N$}
            \label{fig:time_n3}
        \end{subfigure} \hfill
        \begin{subfigure}[t]{0.32\textwidth}
            \centering
            \includegraphics[width = \linewidth]{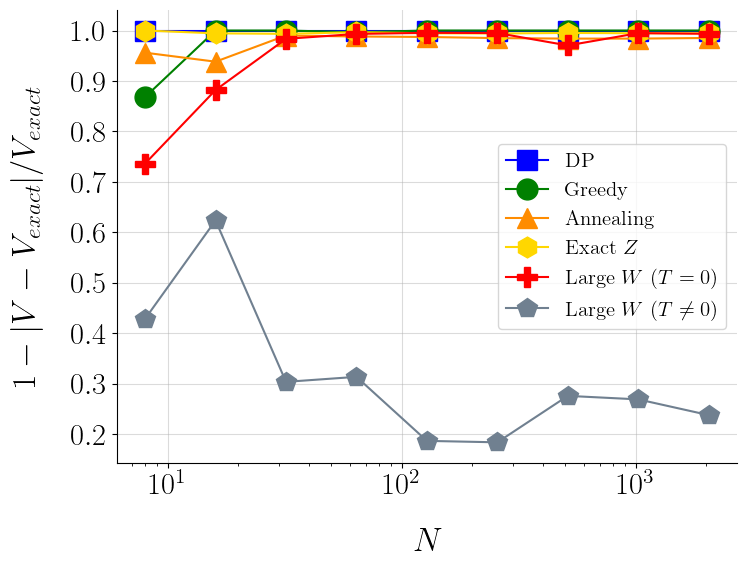}
            \caption{Profit-Ceiling: Acc. vs. $N$}
            \label{fig:v_n3}
        \end{subfigure}\hfill
        \begin{subfigure}[t]{0.32\textwidth}
            \centering
            \includegraphics[width = \linewidth]{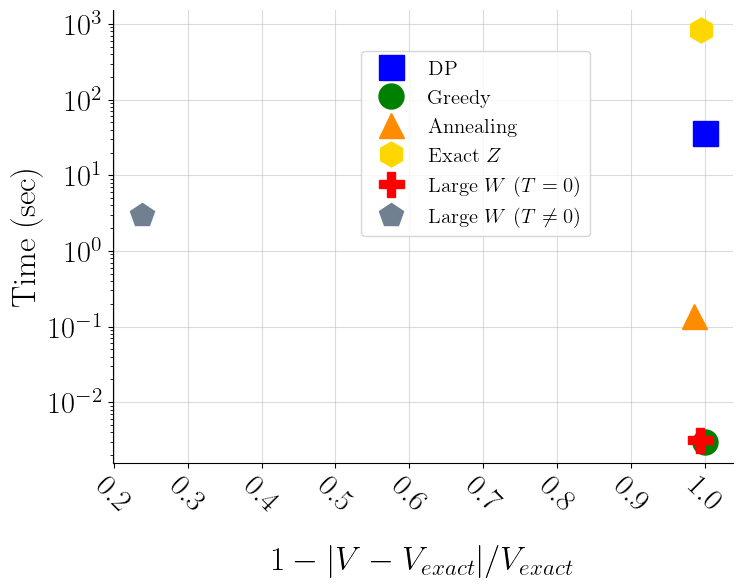}
            \caption{Profit-Ceiling: Time vs. Acc.}
            \label{fig:time_v3}
        \end{subfigure}\\\vspace{.5cm}
        \begin{subfigure}[t]{0.32\textwidth}
            \centering
            \includegraphics[width = \linewidth]{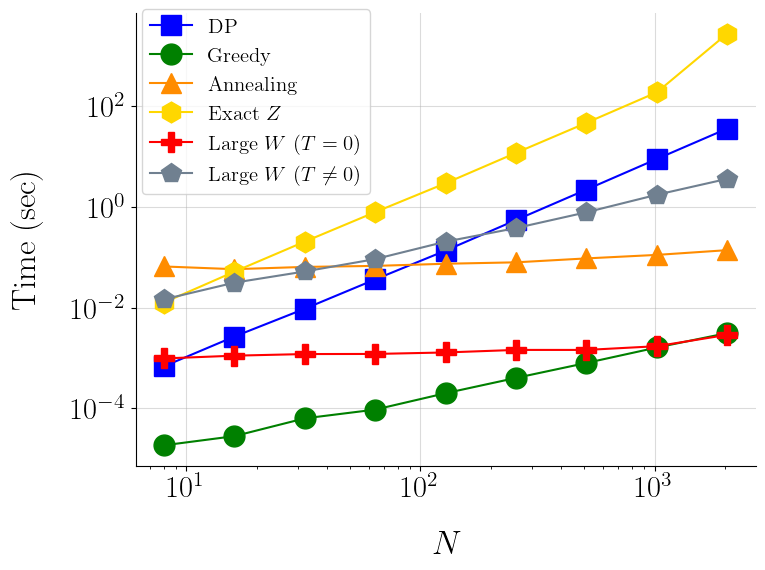}
            \caption{Multi-Strong: Time vs. $N$}
            \label{fig:time_n4}
        \end{subfigure} \hfill
        \begin{subfigure}[t]{0.32\textwidth}
            \centering
            \includegraphics[width = \linewidth]{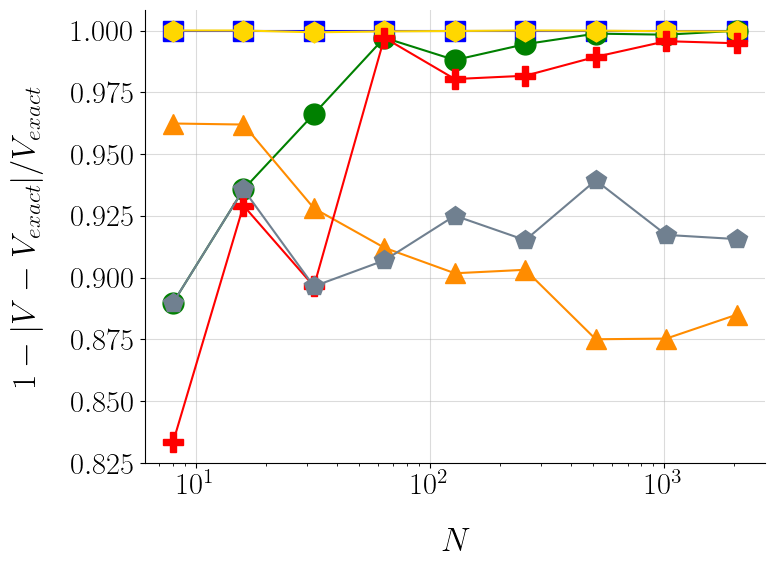}
            \caption{Multi-Strong: Acc. vs $N$}
            \label{fig:v_n4}
        \end{subfigure}\hfill
        \begin{subfigure}[t]{0.32\textwidth}
            \centering
            \includegraphics[width = \linewidth]{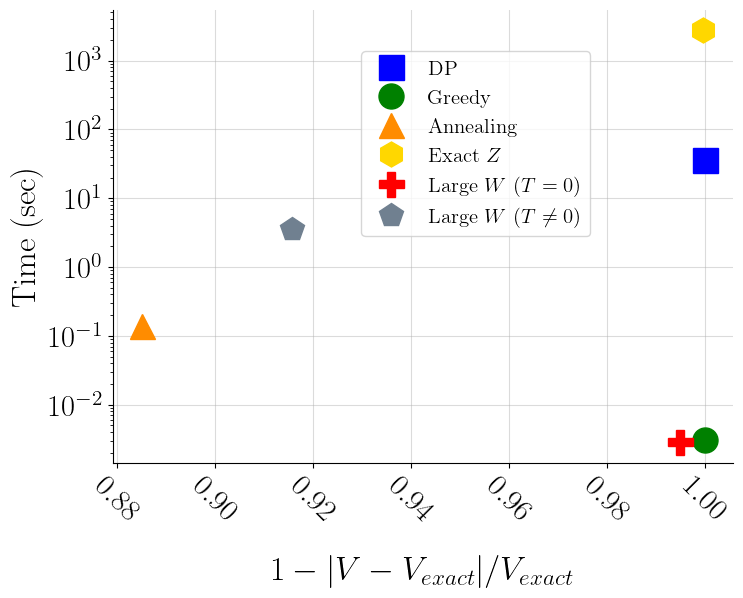}
            \caption{Multi-Strong: Time vs. Acc.}
            \label{fig:time_v4}
        \end{subfigure}    
        \caption{Runtime and Accuracy Comparisons: In (a),(b), and (c), we show the results for the "profit-ceiling instances" and in (d), (e), and (f), we show the results for the "multiple strongly correlated items instances." (a) and (d) depict the runtimes as a function of $N$. (b) and (e) depict the accuracy of the algorithms as a function of $N$. (c) and (f) depict the runtime and accuracy for the instance with $N=2048$ items. For the temperature dependent algorithms we set $T=1.0$. The zero-temperature algorithm has increasing performance with increasing $N$ as is consistent with Fig. \ref{fig:accuracy_complexity}. Code and results used to generate these figures is linked to in Sec. \ref{sec:code}, \textit{Data Availability Statement}. 
        }
        \label{fig:spd_accy2}
    \end{figure*}

    The results of applying the three introduced algorithms and the three standard algorithms to these instances are shown in in \reffig{spd_accy2}. The results are similar to those in \reffig{spd_accy}, except since neither the "multi-strong" or the "profit-ceiling" instances exhibit high degeneracies in their value-to-weight ratios, we find that the zero-temperature algorithm performs better than it did for the "spanner" instance. For both types of instances, the standard greedy algorithm achieves highly accurate results as $N$ increases (due to the $W\gg w_j$ limit), and the zero-temperature algorithm improves as $N$ increases but remains bounded above by the standard greedy algorithm. The nonzero-temperature algorithm fails to achieve accurate results for the profit-ceiling instance, but maintains an accuracy above $90\%$ for the multiple-strongly correlated items instance. \reffig{time_v3} and \reffig{time_v4} show the close proximity in time and accuracy space between the zero-temperature algorithm and the standard greedy algorithm which further motivates identifying the former as a kind of greedy algorithm. In all, these results affirm those in the text and provide additional instances where the introduced algorithms do not perform better in runtime or accuracy than their standard counterparts.




\newcommand{\etalchar}[1]{$^{#1}$}
\newcommand{\noopsort}[1]{} \newcommand{\printfirst}[2]{#1}
  \newcommand{\singleletter}[1]{#1} \newcommand{\switchargs}[2]{#2#1}


\end{document}